\documentclass{ptephy_v1}

\usepackage{amsmath}
\usepackage{amsfonts}
\usepackage{amssymb}
\usepackage{ulem}
\usepackage{eucal}

\usepackage{color}
\usepackage{graphicx}

\definecolor{red  }{rgb}{1,0,0}
\definecolor{blue }{rgb}{0,0,1}
\definecolor{green}{rgb}{0,1,0}
\definecolor{CiteColor}{rgb}{0,0,0.35}
\definecolor{URLColor}{rgb}{0,0,0.35}
\definecolor{darkgreen}{rgb}{0.2,0.7,0.2}

\allowdisplaybreaks
\begin{document}

\title{Multiband Gravitational-Wave Astronomy: 
Observing binary inspirals with a decihertz detector, B-DECIGO}

\author{\name{Soichiro Isoyama}{1},
\name{Hiroyuki Nakano}{2}, 
and
\name{Takashi Nakamura}{3,4}}

\address{${}^1$\affil{1}{International Institute of Physics, 
Universidade Federal do Rio Grande do Norte, Campus Universitario, Lagoa Nova, 
Natal-RN 59078-970, Brazil}
\\
${}^2$\affil{2}{Faculty of Law, Ryukoku University, Kyoto 612-8577, Japan}
\\
${}^3$\affil{3}{Center for Gravitational Physics, Yukawa Institute 
for Theoretical Physics, Kyoto University,Kyoto 606-8502, Japan}
\\
${}^4$\affil{4}{Department of Physics, Kyoto University, 
Kyoto 606-8502, Japan}}


\begin{abstract}
An evolving Japanese gravitational-wave (GW) mission in the deci-Hz band: B-DECIGO (DECihertz laser Interferometer Gravitational wave Observatory) will enable us to detect GW150914-like binary black holes, GW170817-like binary neutron stars, and intermediate-mass binary black holes out to cosmological distances. The B-DECIGO band slots in between the aLIGO-Virgo-KAGRA-IndIGO (hecto-Hz) and LISA (milli-Hz) bands for broader bandwidth; the sources described emit GWs for weeks to years across the multiband to accumulate high signal-to-noise ratios. This suggests the possibility that joint detection would greatly improve the parameter estimation of the binaries. We examine B-DECIGO's ability to measure binary parameters and assess to what extent multiband analysis could improve such measurement. Using non-precessing post-Newtonian waveforms with the Fisher matrix approach, we find for systems like GW150914 and GW170817 that B-DECIGO can measure the mass ratio to within $< 0.1\%$, the individual black-hole spins to within $< 10\%$, and the coalescence time to within $< 5\,$s about a week before alerting aLIGO and electromagnetic facilities. Prior information from B-DECIGO for aLIGO can further reduce the uncertainty in the measurement of, e.g., certain neutron star tidally-induced deformations by factor of $\sim 6$, and potentially determine the spin-induced neutron star quadrupole moment. Joint LISA and B-DECIGO measurement will also be able to recover the masses and spins of intermediate-mass binary black holes at percent-level precision. However, there will be a large systematic bias in these results due to post-Newtonian approximation of exact GW signals.

\end{abstract}

\subjectindex{E02, E31, E32, F31, F33}

\maketitle

\section{Introduction}
\label{sec:intro}

Similar to the electromagnetic spectrum, gravitational waves (GWs) 
have a gravitational spectrum that covers frequencies of GWs ranging 
from $10^{-8}\,{\mathrm {Hz}}$ to $10^{3}\,{\mathrm {Hz}}$, 
broadly divided into the four bands: nano-Hz, milli-Hz, deci-Hz and hecto-Hz. 
GWs in the nano-Hz range are actively sought 
by the Pulsar Timing Array~\cite{Verbiest:2016vem}, 
and the milli-Hz band will be visible by
LISA~\cite{Audley:2017drz,Armano:2018kix} 
(see also Refs.~\cite{Ando:2010zz,Ni:2016wcv}). 
The hecto-Hz band has now been opened up 
by Advanced LIGO (aLIGO)~\cite{TheLIGOScientific:2014jea} 
and Advanced Virgo~\cite{TheVirgo:2014hva} 
with six detections of binary black holes
(BBHs)~\cite{Abbott:2016blz,Abbott:2016nmj,Abbott:2017vtc,
Abbott:2017oio,Abbott:2017gyy}
and binary neutron stars (BNSs)~\cite{TheLIGOScientific:2017qsa}. 
This band will soon be probed more deeply 
by the upcoming KAGRA~\cite{Kuroda:1999vi,Somiya:2011np,Akutsu:2017kpk},  
IndIGO~\cite{Unnikrishnan:2013qwa}, 
and, further into the future, $3$rd generation GW interferometers, 
e.g., Einstein Telescope (ET)~\cite{Hild:2009ns} 
(see also Refs.~\cite{Adhikari:2013kya,Miller:2014kma,Dwyer:2014fpa}).
The remaining deci-Hz band is the target of 
\textit{the DECihertz laser Interferometer Gravitational wave
Observatory} (B-DECIGO): a planned Japanese
space-borne detector~\cite{Nakamura:2016hna}
\footnote{Pre-DECIGO in Ref.~\cite{Nakamura:2016hna} is the same as B-DECIGO. 
After publication of Ref.~\cite{Nakamura:2016hna}, 
the DECIGO Working Group decided to change the name from ``Pre-DECIGO''  
to ``B-DECIGO'' because there is a strong science case for B-DECIGO 
and ``Pre'' is not the best word to describe the mission.}.
The original DECIGO mission concept was proposed 
by Seto, Kawamura, and Nakamura in 2001~\cite{Seto:2001qf}, 
and B-DECIGO is the scaled-down version of DEICGO set to be 
the ``$1$st-generation'' of the deci-Hz GW detectors 
(B-DECIGO stands for ``Basic'' or ``Base'' DECIGO). 
B-DECIGO will consist of three satellites in a $100~{\mathrm {km}}$ 
equilateral triangle, having sun-synchronous dusk-dawn circular orbits 
$2000~{\mathrm {km}}$ above the Earth~\cite{Sato:2017dkf,Musha:2017usi}
\footnote{The original DECIGO mission concept consists of three satellites 
with a $1000~{\mathrm {km}}$ equilateral triangle heliocentric orbit 
and four similar systems. This will allow the precise measurement 
of the direction and polarization of GW sources~\cite{Kawamura:2011zz}.}.
With B-DECIGO operating, we will probe this deci-Hz window 
for the first time, completing the full gravitational spectrum;  
see FIG.~\ref{fig:PSD-SNR}.

Compact binary coalescence is a key target in GW astronomy, 
and B-DECIGO has two clear targets to observe. 
The first \textit{promised} targets are the inspiraling 
GW150914-like BBHs and GW170817-like
BNSs~\cite{Abbott:2016blz,Abbott:2016nmj,Abbott:2017vtc,
Abbott:2017oio,Abbott:2017gyy}.
The second \textit{original} targets will be the merger 
of intermediate-mass BBHs with total mass between a few hundreds and 
$\sim 10^{4} M_{\odot}$~\cite{Miller:2003sc,Gurkan:2005xz,Abbott:2017iws}
\footnote{The event rate of intermediate-mass BBHs depends 
on the astrophysical models that we assume 
(see e.g., Ref.~\cite{Yagi:2012gb,Mandel:2017pzd}). 
We do not attempt an event rate estimation here.}. 
A key goal of the B-DECIGO project is to explore 
the origin and evolution history of these BNSs and BBHs, 
precisely measuring their parameters 
(masses, spins, Love numbers etc.) out to the high-redshift universe
~\cite{Kinugawa:2014zha,Nakamura:2016hna,Chen:2017gfm,Farr:2017uvj} 
(other scientific cases for deci-Hz detectors are summarized 
in, e.g., Refs.~\cite{Crowder:2005nr,Kawamura:2011zz,Mandel:2017pzd}).

Besides opening the deci-Hz window to GW astronomy, 
a novelty of B-DECIGO detection lies in the fact that 
it will be followed up and will guide the binary measurement in other bands. 
In FIG.~\ref{fig:PSD-SNR}, 
we immediately see that the early inspiral parts 
of binary systems like GW150914 and GW170817 are visible
in the deci-Hz (B-DECIGO) band 
(and even in the milli-Hz (LISA) band for GW150914) 
prior to coalescence in the hecto-Hz (aLIGO) band. 
The binary parameters that are most readily accessible differ in each band,  
which, in turn, suggests that the joint multiband analysis 
across LISA, B-DECIGO, and aLIGO bands may be able to 
greatly enhance our ability to measure compact binaries.

The above observation has motivated the formulation of 
\textit{multiband GW astronomy} over the full gravitational spectrum. 
Shortly after the first GW150914 detection, 
Sesana underlined the concept of multiband GW astronomy 
with aLIGO and LISA~\cite{Sesana:2016ljz}.  
It is now realized that LISA will be able to measure 
GW150914-like BBHs up to some thousands of them
in the low-redshift universe~\cite{Sesana:2016ljz,Kyutoku:2016ppx}. 
Such measurement will accurately determine their sky position 
for aLIGO~\cite{Sesana:2016ljz}, distinguish their formation channels
~\cite{Seto:2016wom,Nishizawa:2016jji,Breivik:2016ddj,Inayoshi:2017hgw},
and provide a new class of cosmological standard
sirens~\cite{Kyutoku:2016zxn,DelPozzo:2017kme}.  
In another development, multiband GW astronomy has been 
investigated for parameter estimation and gravity testing. 
Nair, Jhingan, and Tanaka have explored these aspects 
(as well as the sky-localization of the sources), 
focusing on the joint DECIGO and ground-based observation 
of BBHs and BH-NS binaries~\cite{Nair:2015bga,Nair:2018bxj}. 
The joint LISA and aLIGO analysis of GW150914-like BBHs 
will also improve the constraints on their 
dipole radiation~\cite{Barausse:2016eii}, 
the uncertainties in parameter estimation, 
and tests of general relativity~\cite{Vitale:2016rfr}. 
Other proposals for multiband GW astronomy
include those in Refs.~\cite{Sesana:2009wg,AmaroSeoane:2009ui,Kocsis:2011jy,
Yagi:2012gb,Yagi:2013du}.

Nakamura et al.~\cite{Nakamura:2016hna} have initiated examination of 
the precision with which the binary parameters can be determined by B-DECIGO, 
considering GW150914-like non-spinning BBHs. 
However, the measurability of BH spins, GW170817-like BNSs 
and the prospects for the multiband analysis 
with B-DECIGO have not yet been fully revealed. 
We shall assess how precisely we are able to measure 
the parameters of BNS and BBH inspirals with B-DECIGO,  
and explore how multiband B-DECIGO and aLIGO/ET (or LISA) 
measurement improves their parameter estimation 
and science cases over those using only single-band detection.
Some previous studies of parameter estimation
in the DECIGO mission can be found
in Refs.~\cite{Takahashi:2003wm,Yagi:2009zz,Yagi:2011wg,
Yagi:2012gb,Yagi:2013du,Nair:2015bga}.

\subsection{Observable range of B-DECIGO}

To set the stage, we first review B-DECIGO's observable range 
for a given detection threshold of the matched-filter 
signal-to-noise ratios (SNRs; see Eq.~\eqref{SNR-ave})~\cite{Nakamura:2016hna}.
Consider binary systems with component masses $m_{1,2}$ 
(we assume $m_1 < m_2$), total mass $m \equiv m_1 + m_2$, 
symmetric mass ratio $\nu \equiv m_1 m_2 / m^2$,  
and chirp mass ${\cal M} \equiv \nu^{3/5} m$ 
(throughout, we use geometric units, where $G=c=1$,
with the useful conversion factor $1 M_{\odot} = 1.477 \; 
{\mathrm {km}} = 4.926 \times 10^{-6} \; {\mathrm{s}}$). 

Assuming quasi-circular inspiraling binaries, 
the coalescing time $t_c$, 
the instantaneous number of GW cycles $N_c \equiv f^2 (df/dt)^{-1}$ 
with the GW frequency $f$, 
and the dimensionless characteristic strain amplitude $h_c$ 
of the binary are estimated as 
(normalized to GW150914-like BBHs;
see Sec.~\ref{sec:result})~\cite{Berti:2004bd,Moore:2014lga} 
\begin{align}\label{estimate-t}
t_c &= 1.03 \times 10^6 \,{\mathrm {s}} \,
\left(\frac{{\cal M}}{30.1 M_{\odot}} \right)^{-5/3} 
\left(\frac{f}{0.1 \,{\mathrm {Hz}}} \right)^{-8/3} \,, \\
\label{estimate-N}
N_c &= 2.75 \times 10^5  
\left(\frac{{\cal M}}{30.1 M_{\odot}} \right)^{-5/3} 
\left(\frac{f}{0.1 \,{\mathrm {Hz}}} \right)^{-5/3} \,, \\
\label{estimate-h}
h_c &= 3.91 \times 10^{-21} 
\left(\frac{{\cal M}}{30.1 M_{\odot}} \right)^{5/6} 
\left(\frac{D_{\mathrm {L}}}{0.4 \,{\mathrm {Gpc}}} \right)^{-1} 
\left(\frac{f}{0.1 \,{\mathrm {Hz}}} \right)^{-1/6} \,,
\end{align}
where $D_{\mathrm {L}}$ is the luminosity distance 
between the observer and the binary, 
and the chirp mass ${\cal M}$ as well as GW frequency $f$ are measured 
\textit{at the observer}, accounting for cosmological effects. 
When binaries are at a cosmological distance, in the geometrical units, 
all mass scales are redshifted by a Doppler factor of $1 + z$ 
with the source's cosmological redshift $z$. 
As a result, for instance, the GW frequency $f^S$ and component masses 
$m_i^S$ at the source location are related to those 
at the observer $f^O$ and $m_i^O$ 
via $f^O = f^S (1 + z)^{-1}$ and $m_i^O = m_i^S (1 + z)$, respectively. 
In the rest of this paper, we always quote physical quantities 
measured at the observer, dropping the labels `$S$' and `$O$' 
(unless otherwise specified)
\footnote{When we need to convert between the source redshift $z$ 
and luminosity distance $D_{\mathrm {L}}$,  
we adopt the Planck flat cosmology~\cite{Ade:2015xua} 
with the matter density parameter ($\Omega_M = 0.31$), 
the dark energy density parameter ($\Omega_\Lambda = 0.69$), 
and the Hubble constant $H_0 = 67.7 \,{\mathrm {km\, s^{-1}\, Mpc^{-1}}}$.
The luminosity distance as a function of redshift $z$ is then given by 
$D_{\mathrm {L}}(z) = ({(1 + z)}/{H_0})
\int_0^z {d z'}{({\Omega_M (1 + z')^3 + \Omega_{\Lambda}})^{-1/2}}$.
}.

Equations~\eqref{estimate-t} -~\eqref{estimate-h} show that 
in the B-DECIGO band GW150914 and GW170817 were visible 
for $\sim 10~{\mathrm {days}}$ and $ \sim 7 ~{\mathrm {yrs}}$ prior 
to coalescence 
with large numbers of GW cycles of $\sim 10^5$ and $\sim 10^7$, respectively.  
More importantly, both of their characteristic strains 
in this deci-Hz band are $h_c \sim 10^{-21}$, 
which are well above the target dimensionless noise amplitude 
of B-DECIGO $\sim 10^{-23}$ around $1~{\mathrm {Hz}}$~\cite{Nakamura:2016hna}. 
This allows B-DECIGO to observe GW150914- and GW170817-like binary inspirals 
out to $\sim 60~{\mathrm {Gpc}}$ ($z \sim 6$)
and $\sim 1~{\mathrm {Gpc}}$ ($z \sim 0.2$), respectively, 
assuming a detection (sky and polarization averaged) SNR threshold at $8$ 
for a $4~{\mathrm {yr}}$ mission lifetime
\footnote{In practice, the detection SNR threshold is set not only 
by the false alarm rate, but also by the computational burden of 
generating inspiral templates. Because of this computational limitations, 
the actual SNR threshold for BNSs would have to be higher value than
$8$~\cite{Cutler:2005qq}.}.

\begin{figure}[tbp]
\begin{tabular}{cc}  
\begin{minipage}[t]{.47\hsize}
 \centering
 \includegraphics[width=\columnwidth, bb = 2 44 706 522]{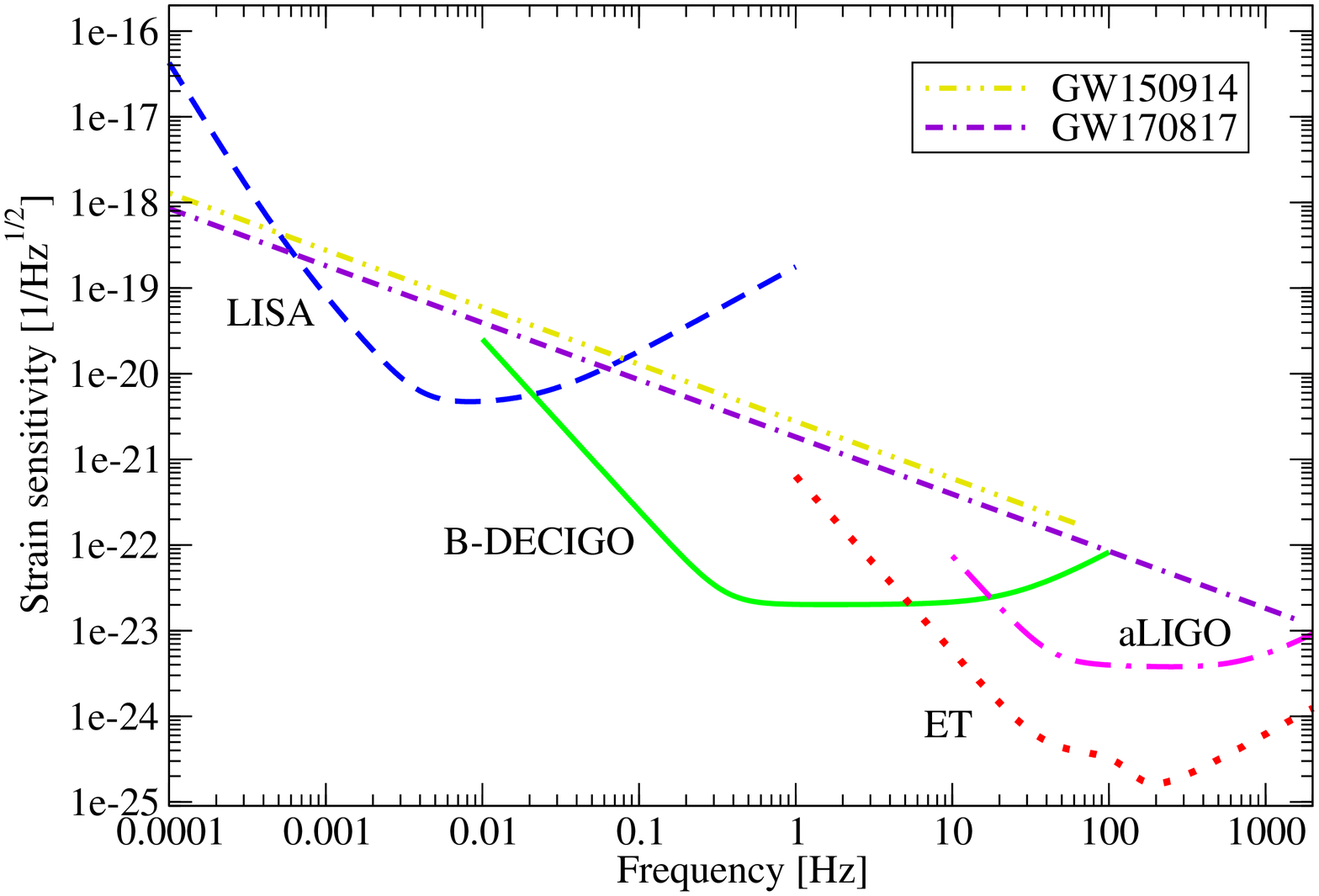}
\end{minipage}
\qquad 
\begin{minipage}[t]{.47\hsize}
 \centering
 \includegraphics[width=\columnwidth, bb = 16 40 732 522]{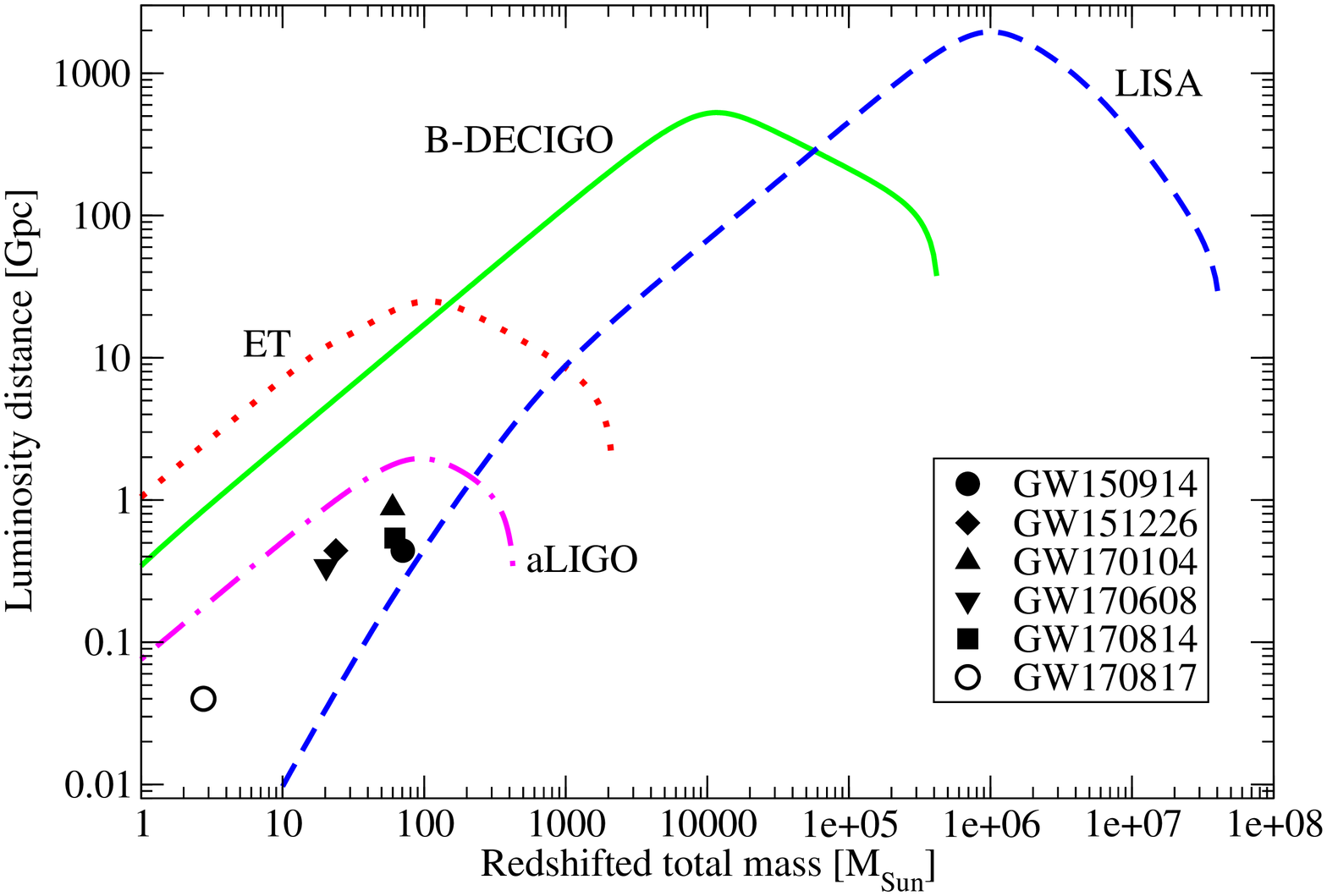}
  \end{minipage}
\end{tabular}
\caption{\textit{Left}: The square root of the noise power spectrum density 
of B-DECIGO, as well as aLIGO, the Einstein Telescope (ET), 
and LISA against GW frequencies (see Sec.~\ref{subsec:PSD}).
The strain sensitivity $h_c(f) f^{-1/2}$~\cite{Moore:2014lga}
completed by GW150914 and GW170817 (in the inspiral phase) 
are also depicted as references. 
\textit{Right}: The detectable luminosity distance $R_{\mathrm {L}}(m)$ 
for equal-mass inspirals as a function of their redshifted total masses $m$. 
We assume a four-year mission lifetime 
for B-DECIGO and LISA~\cite{Audley:2017drz} 
and a detection SNR threshold at $8$,  
for which we average over all-sky positions and binary orientation 
(see Eq.~\eqref{SNR-ave}); $R_{\mathrm {L}}(m)$ with $m$ fixed becomes smaller 
by a factor of $\sqrt{4 \nu}$ for unequal-mass systems 
and larger by a factor of $2.5$ for the optimal geometry.
The luminosity distances of BBHs and a BNS observed by aLIGO and Virgo 
are also marked. These figures compliment the similar 
Figs.~2 and 3 in Ref.~\cite{Nakamura:2016hna}.}
\label{fig:PSD-SNR}
\end{figure}

Intermediate-mass BBHs with the redshifted total mass 
$m \sim 10^4 M_{\odot}$ can also stay in the B-DECIGO band 
for $\sim  1~{\mathrm {d}}$ with $\sim 10^3$ GW cycles
before their final merger. 
Indeed, the GW frequency emitted at the innermost stable circular orbit 
(ISCO) of a Schwarzschild BH with (redshifted) mass $M$ is 
\begin{equation}\label{f-ISCO}
f_{\mathrm {ISCO}} = 
0.44 ~{\mathrm {Hz}} \, \left(\frac{M}{10^4 M_{\odot}} \right)^{-1}\,,
\end{equation}
placing GWs emitted from such intermediate-mass BBHs 
well within the B-DECIGO band. 
Once again assuming a detection (sky and polarization averaged) 
SNR threshold of $8$, we see that they are within the observable range 
of B-DECIGO even at $\sim 520\,{\mathrm {Gpc}}$ ($z \sim 40$).  
These results are illustrated in FIG.~\ref{fig:PSD-SNR}, 
with the conclusion that the high-redshift BNSs and BBHs 
up to $m \sim 10^5 M_{\odot}$ can indeed be detectable by B-DECIGO.

\subsection{Parameter estimation with B-DECIGO: Multiband measurement}

In matched filtering analysis, the measurement errors of binary parameters 
are classified into two categories; the statistical error 
due to the random noise in the detectors, and the systematic errors 
resulting from, e.g., inaccurate waveform modeling 
of the expected GW signals. 
To underline what could be expected from B-DECIGO measurements, 
next we consider the inspiral phase of aligned-spin 
BNSs and BBHs in quasi-circular orbits, 
which is approximated well by the post-Newtonian (PN)
waveforms~\cite{Blanchet:2013haa}.

For B-DECIGO, the achievable precision of the statistical error 
can be very high, because the statistical errors scale as 
$\propto {\mathrm {SNR}}^{-1}
(1 + O({\mathrm {SNR^{-1}}}))$~\cite{Vallisneri:2007ev,Rodriguez:2013mla}, 
and FIG.~\ref{fig:PSD-SNR} shows that BNSs and BBHs 
will be visible by B-DECIGO with high SNRs
\footnote{For an inspiral with the redshifted total mass $m$ 
at the luminosity distance $D_{\mathrm {L}}$, 
the (sky and polarization averaged) SNR is estimated as 
$8 (D_{\mathrm {L}} / R_{\mathrm {L}}(m))^{-1}$, making use of the detectable luminosity distance 
$R_{\mathrm {L}}(m)$ in FIG.~\ref{fig:PSD-SNR}.}.
At the same time, recall that the precision in statistical errors 
is determined by a combination of the SNR and the \textit{bandwidth}  
over which a detector accumulates the SNR 
(see, e.g., Ref.~\cite{Arun:2004hn}). 
Indeed, the high SNR of systems like GW150914 and GW170817 
in the B-DECIGO band comes from the large numbers of GW cycles of $\sim 10^6$ 
(recall Eq.~\eqref{estimate-N}) accumulated in the much greater bandwidth 
during inspiral than that in the aLIGO band. 
This suggests the interesting possibility that 
we may be able to precisely measure, for instance, 
individual BH spins in GW150914-like BBHs in the B-DECIGO band, 
which can be hard to measure in the aLIGO band 
because of their strong degeneracy in the parameter dependence of the PN
waveform~\cite{Cutler:1994ys,Poisson:1995ef,Baird:2012cu,Hannam:2013uu,
Ohme:2013nsa,Purrer:2015nkh}. 
The effects of BH spins come into the waveform at higher frequencies, 
but the broader bandwidth available in the lower B-DECIGO band 
might allow for tighter constraints. 

The multiband analysis with B-DECIGO will further 
reduce the statistical uncertainties using only single-band analysis. 
The key point is that information from earlier B-DECIGO (or LISA) analysis 
can constrain the prior on the aLIGO (or B-DECIGO) analysis most naturally. 
This prior information refines the estimation of, for instance, 
the NS spin- and tidally induced deformations in BNS waveforms, 
from which we can infer the NS internal structure 
(equation of state)~\cite{Abbott:2018wiz,Abbott:2018exr}.
Once again, their precision is limited by a partial degeneracy 
between these effects and the mass ratios, as well as spins 
in the PN waveform~\cite{Poisson:1997ha,Flanagan:2007ix,
Favata:2013rwa,Agathos:2015uaa,Harry:2018hke,Dietrich:2018uni}.
If the mass ratio is already precisely constrained  
from the early inspiral in the lower B-DECIGO band, 
the measurement of such matter imprints in the waveform 
could be more precise in the higher aLIGO band.

Nonetheless, these improvement in statistical errors  
could be hampered by the systematic errors 
in the PN modeling of BNS and BBH inspirals, 
which is still an approximation of true general-relativistic 
(numerical-relativity) waveforms. 
Here it is important to recognize that the systematic measurement error 
is \textit{SNR independent}~\cite{Cutler:2007mi} 
while the statistical measurement error scales with the inverse of SNR. 
Because the BNSs and BBHs observable by B-DECIGO will have high SNR, 
ultimately, the systematic mismodeling errors might be the limiting factor 
in their measurement. 


The remainder of the paper quantifies the statistical and systematic 
parameter estimation errors in inspiraling BNSs and BBHs, 
using B-DECIGO and multiband measurements. 
We begin in Sec.~\ref{sec:GWs} with a review of our methodology, 
including our PN waveform model and the Fisher matrix formalism 
for parameter estimation of GW signals. 
In Sec.~\ref{sec:result}, we present our main results 
for the statistical and systematic errors, respectively. 
We finish with two scientific cases in Sec.~\ref{sec:discuss} 
that could be done in multiband GW astronomy with B-DECIGO: 
the redshift measurement of cosmological BNSs 
with GW observation alone, 
and the characterization of final remnant BHs using BBH inspirals.

\section{Parameter estimation using post-Newtonian waveforms}
\label{sec:GWs}

As a first step toward the full problem of parameter estimation, 
for simplicity, we use as our GW signal model 
the up-to-date inspiral-only PN waveform with tidal (finite-size)
corrections~\cite{Blanchet:2013haa,Flanagan:2007ix,Alvi:2001mx} 
neglecting merger and ringdown, 
and employ the semi-analytic Fisher information matrix formalism
~\cite{Finn:1992wt,Cutler:1994ys,Cutler:2007mi} for the signal analysis. 
We will also neglect the orbital motion of B-DECIGO and LISA, and 
average the signals over the all-sky positions and binary orientations
~\cite{Finn:1992wt,Dalal:2006qt}. 
Our analysis is limited to the 
aligned-spin inspirals in quasi-circular orbits 
because these are the only configurations 
for which the NS and BH tidal influences on the GW phase 
are computed in the PN approximation.

\subsection{Tidally corrected non-precessing $3.5$PN waveform} 
\label{subsec:PN}

Motivated by the fact that matched filtering is more sensitive 
to the phase of the signal than its amplitude, 
we work with the frequency-domain ``restricted'' stationary 
phase approximation to the PN waveform, in which 
both higher-multipole components 
and PN corrections to the wave amplitude are ignored. 
After averaging over the all-sky positions and binary orientations, 
the resultant waveform reads~\cite{Berti:2004bd}  
\begin{equation}\label{hf}
{\tilde h}(f) = {\cal A} f^{-7/6}  e^{i \Psi(f) }\,, 
\quad
{\cal A} = 
\frac{2}{5} \times  \sqrt{\frac{5}{24}} \pi^{-2/3} 
\frac{{\cal M}^{5/6}}{D_{\mathrm {L}}}\,, 
\end{equation}
with the ``Newtonian'' amplitude scaled by the factor $2/5$. 
The GW phase $\Psi(f)$ is the sum of two contributions: 
(i) spinning point-particle terms 
that are independent of the nature of the NSs or BHs 
comprising the binary, superposed with (ii) finite-size terms 
(depending on the nature of the NS or BH comprising the binary) 
that arise from the rotational deformation of an axially symmetric NS 
\footnote{Following the tradition in PN waveforms, 
we have classified the corrections from BH rotational deformation here 
as ``spinning point-particle'' terms~\cite{Arun:2008kb,Mishra:2016whh}.}, 
and the tidal response of the BH or NS in the binary 
on the other companion.

For non-precessing BNSs, the GW phase $\Psi(f)$ may be expressed as 
\begin{align}\label{F2-BNS}
\Psi_{\mathrm {BNS}}(f)
&=   
2 \pi f t_c - \Psi_c - \frac{\pi}{4} \cr 
& \quad +
\frac{3}{128 \nu v^5}
\left( 
\Delta \Psi^{\mathrm {pp}}_{3.5{\mathrm {PN}}}
+
\Delta \Psi^{\mathrm {pp-spin}}_{3.5{\mathrm {PN}}}
+
\Delta \Psi^{\mathrm {NS-QM}}_{3.5{\mathrm {PN}}}
+
\Delta \Psi^{\mathrm {NS-tidal}}_{6 {\mathrm {PN}}}
\right)\,, 
\end{align}
where $t_c$ and $\Psi_c$ are the coalescence time and phase, respectively, 
and we introduce the orbital velocity $v \equiv ( \pi m f )^{1/3}$: 
an $O(v^{2n})$ term is of $n$th PN order.
The first term, $\Delta \Psi^{\mathrm {pp}}_{3.5{\mathrm {PN}}}$, 
is the spin-independent, point-particle contribution up to $3.5$PN order, 
derived in Ref.~\cite{Arun:2004hn}
\footnote{In our analysis, we break with tradition and 
keep only terms $ \propto \ln(v)$ at $2.5$PN order. 
The terms $\propto v^5$ become constant 
in $\Psi_{\mathrm {BNS}}(f)$ due to cancellation 
with the overall factor of $v^{-5}$, 
and can be absorbed into $\Psi_c$~\cite{Arun:2004hn}. 
The same is applied with the $2.5$PN terms 
in $\Psi_{{\mathrm {BBH}}}(f)$ below.}. 
The second term, $\Delta \Psi^{\mathrm {pp-spin}}_{3.5{\mathrm {PN}}}$, 
is the $3.5$PN spin-dependent, point-particle contribution 
that includes linear spin-orbit effects~\cite{Blanchet:2013haa,Bohe:2013cla}, 
quadratic-in-spin effects~\cite{Bohe:2015ana}, 
and cubic-in-spin effects~\cite{Marsat:2014xea}.
Using the dimensionless spin parameter 
${\chi}_i \equiv {{\bf{S}}_i^S \cdot {\boldsymbol {\ell}^S}}/{(m_i^S)^2}$ 
defined in terms of the source-frame individual body's mass $m_i^S$ 
and spin vectors ${\bf{S}}_i^S$ as well as the unit normal 
${\boldsymbol \ell}^S$ to the orbital plane, 
$\Delta \Psi^{\mathrm {pp-spin}}_{3.5{\mathrm {PN}}}$ is given 
as a function of $v$ and $(\nu, \chi_s, \chi_a)$ 
where $\chi_s \equiv {(\chi_1 + \chi_2)}/{2}$ and 
$\chi_a \equiv {(\chi_1 - \chi_2)}/{2}$. 
Note that positive (negative) values of ${\chi}_i$ correspond 
to the aligned (anti-aligned) configurations 
with respect to the orbital angular momentum of the binary. 
Their explicit expressions are computed 
in Refs.~\cite{Arun:2008kb,Mishra:2016whh}.

The third term, $\Delta \Psi^{\mathrm {NS-QM}}_{3.5{\mathrm {PN}}}$, 
is the finite-size correction due to the rotational deformation of a NS. 
Restricted to the dominant effect, this is well characterized 
by the (dimensionless) NS quadrupole parameter 
$\kappa_{i} \equiv - (Q / \chi_i^2)/(m_{i}^S)^3$
~\cite{Poisson:1997ha,Bohe:2015ana}
\footnote{The known NS has at most
$\chi \lesssim 0.4$~\cite{Burgay:2003jj,Hessels:2006ze}, 
for which this characterization is sufficient~\cite{Laarakkers:1997hb}.}. 
The spin-induced quadrupole moment scalar $Q$~\cite{Pappas:2012ns} 
is fixed when the source-frame NS mass $m^S$ 
and equation of state are given, encoding the NS internal structure.  
Such spin-induced quadrupole-moment corrections to the GW phase start 
from $2$PN order beyond the lowest PN term in Eq.~\eqref{F2-BNS}, 
and we include them to the $3.5$PN order 
~\cite{Krishnendu:2017shb} (assuming $m_1 < m_2$)
\footnote{We neglect the subdominant spin-induced NS octpole moment 
also entering $3.5$PN order~\cite{Marsat:2014xea,Banihashemi:2018xfb}.}: 
\begin{align}\label{NS-QM}
\Delta \Psi^{\mathrm {NS-QM}}_{3.5{\mathrm {PN}}}
&\equiv 
-25{\tilde {\cal Q}} v^{4}
+
\left\{
\left(\frac{15635}{42} + 60 \nu \right){\tilde {\cal Q}} 
-
\frac{2215}{24} \sqrt{1 - 4 \nu} {\delta {\tilde {\cal Q}}}
\right\} v^{6} \cr
& \quad 
+
\left[
\left\{
-\left(280 \pi + 10 \nu \chi_s \right)
+ 
\frac{375}{2}
\left(\chi_s - \sqrt{1 - 4 \nu} \chi_a \right)
\right\}{\tilde {\cal Q}} \right. \cr 
& \quad \left. 
+
\frac{1985}{6} \left( \chi_a - \sqrt{1 - 4 \nu} \chi_s \right) 
{\delta {\tilde {\cal Q}}}
\right] v^{7} \,,
\end{align}
where we define the ``combined'' dimensionless quadrupole parameters 
scaling as the square of the NS spins by 
\begin{align}\label{def-tQ}
{\tilde {\cal Q}} 
&\equiv
\left\{ 
(1 - 2 \nu) ({\kappa_1 + \kappa_2} - 2) 
- 
\sqrt{1 - 4 \nu} (\kappa_1 - \kappa_2)  
\right\} \left(\chi_s^2 + \chi_a^2\right) \cr
& \quad 
+ 
2 \left\{ 
(1 - 2 \nu)(\kappa_1 - \kappa_2) 
- 
\sqrt{1 - 4 \nu} ({\kappa_1 + \kappa_2} - 2) 
\right\} \chi_s \chi_a\,, \cr 
\delta {\tilde {\cal Q}} 
&\equiv
\left\{ 
(1 - 2 \nu) ({\kappa_1 - \kappa_2}) 
- 
\sqrt{1 - 4 \nu} (\kappa_1 + \kappa_2 - 2)  
\right\} \left(\chi_s^2 + \chi_a^2\right) \cr
& \quad 
+ 
2 \left\{ 
(1 - 2 \nu)({\kappa_1 + \kappa_2} - 2) 
- 
\sqrt{1 - 4 \nu} ({\kappa_1 - \kappa_2}) 
\right\} \chi_s \chi_a\,.
\end{align}
These parameters are conveniently chosen such that 
(a) the leading-order correction at $2$PN order depends 
only on ${\tilde {\cal Q}}$, 
(b) ${\tilde {\cal Q}} = 0 = \delta {\tilde {\cal Q}}$ 
for a BBH because a spinning BH has $\kappa_i = 1$~\cite{Bohe:2015ana}, 
and (c) ${\tilde {\cal Q}} (\kappa_1 = \kappa_2 = \kappa) 
= (\kappa/2 - 1) \chi^2$ as well as 
$\delta {\tilde {\cal Q}} (\kappa_1 = \kappa_2 = \kappa) = 0$
for equal-mass ($\nu = 1/4$), equal-spin ($\chi_1 = \chi_2 = \chi$) BNSs. 
It is also important to recognize that the parameters 
$({\tilde {\cal Q}},\, \delta {\tilde {\cal Q}})$ implicitly 
include the redshift factor $(1 + z)^3$ 
when using the NS masses at the observer $m_i^O$~\cite{Harry:2018hke}.

The last term, $\Delta \Psi^{\mathrm {NS-tidal}}_{6 {\mathrm {PN}}}$, 
is the finite-size correction due to the quadrupole tidal response of a NS. 
Restricted to slowly changing tidal fields, 
this response can be characterized 
by the (dimensionless) NS tidal deformability 
$\Lambda_{i} \equiv (2/3) k_2 (R_{i}^S / m_{i}^S)^5$~\cite{Flanagan:2007ix} 
(see also Ref.~\cite{Gralla:2017djj}). 
Similar to the quadrupole moment scalar $Q$, 
both the second (electric-type) Love number $k_2$~\cite{Binnington:2009bb}
and the source-frame NS radius $R^S$ are fixed 
when $m^S$ and the equation of state are given. 
Such tidal correction to the GW phase starts from 
$5$PN order beyond the lowest PN term in Eq.~\eqref{F2-BNS}, 
and we are only concerned with the leading-order ($5$PN) 
and next-to-leading-order contributions ($6$PN)
~\cite{Vines:2011ud,Damour:2012yf} 
\footnote{The spin-tidal coupling term starts 
at $6.5$PN~\cite{Abdelsalhin:2018reg,Landry:2018bil}, 
which is negligible here.}:
\begin{equation}\label{NS-tidal}
\Delta \Psi^{\mathrm {NS-tidal}}_{6{\mathrm {PN}}} 
\equiv 
-\frac{39}{2}{\tilde \Lambda} v^{10}
+
\left(
-\frac{3115}{64}{\tilde \Lambda} 
+\frac{6595}{364} \sqrt{1 - 4 \nu} {\delta {\tilde \Lambda}}
\right) v^{12}\,,
\end{equation}
where ``combined'' dimensionless tidal deformabilities 
$\tilde \Lambda$ and $\delta {\tilde \Lambda}$ are 
given by (once again assuming $m_1 < m_2$) 
~\cite{Favata:2013rwa,Wade:2014vqa,TheLIGOScientific:2017qsa} 
\begin{align}\label{def-tL}
{\tilde \Lambda}
&\equiv
\frac{8}{13}
\left\{
(1 + 7 \nu - 31 \nu^2) (\Lambda_1 + \Lambda_2) 
-
\sqrt{1 - 4 \nu} (1 + 9 \nu - 11 \nu^2) (\Lambda_1 - \Lambda_2) 
\right\}\,, \cr
\delta {\tilde \Lambda}
&\equiv
\frac{1}{2}
\left\{
\sqrt{1 - 4 \nu}
\left(1 - \frac{13272}{1319} \nu - \frac{8944}{1319} \nu^2 \right) 
(\Lambda_1 + \Lambda_2) 
\right. \cr
& \quad \left. 
-
\left( 
1 - \frac{15910}{1319} \nu + \frac{32850}{1319} \nu^2 
+ \frac{3380}{1319} \nu^3 
\right) (\Lambda_1 - \Lambda_2) 
\right\}\,.
\end{align}
They have the convenient properties 
$\tilde \Lambda (\Lambda_1 = \Lambda_2 = \Lambda) = \Lambda$ and 
$\delta {\tilde \Lambda} (\Lambda_1 = \Lambda_2 = \Lambda) = 0$ 
in the equal-mass limit $\nu = 1/4$.
The parameters $\tilde \Lambda$ and $\delta {\tilde \Lambda}$ 
in Eq.~\eqref{NS-tidal} are redefined 
to include the Doppler factor $(1 + z)^5$ 
for $\Lambda_{1,2}$ because we have used the NS masses 
at the observer $m_i^O$.  
We also remark that each  NS tidal deformability can be 
related to the NS's spin-induced quadrupole-moment parameter 
using quasi-universal relations~\cite{Yagi:2016bkt}.

Meanwhile, the GW phase $\Psi(f)$ for aligned-spin BBHs may have the form 
\begin{align}\label{F2-BBH}
\Psi_{\mathrm {BBH}}(f)
&=   
2 \pi f t_c - \Psi_c - \frac{\pi}{4} 
+
\frac{3}{128 \nu v^5}
\left( 
\Delta \Psi^{\mathrm {pp}}_{3.5{\mathrm {PN}}}
+
\Delta \Psi^{\mathrm {pp-spin}}_{3.5{\mathrm {PN}}}
+
\Delta \Psi^{\mathrm {BH-tidal}}_{3.5{\mathrm {PN}}} 
\right)\,.
\end{align}
The spin-(in)dependent point-particle terms 
$\Delta \Psi^{\mathrm {pp}}_{3.5{\mathrm {PN}}}$ 
and $\Delta \Psi^{\mathrm {pp-spin}}_{3.5{\mathrm {PN}}}$ 
are the same as those for a BNS in Eq.~\eqref{F2-BNS}. 
The third term, $\Delta \Psi^{\mathrm {BH-tidal}}_{3.5{\mathrm {PN}}}$,  
is the finite-size correction due to the tidal response of a BH. 
Restricted to slowly changing tidal fields, 
each BH in a BBH is tidally heated and torqued by its 
companion~\cite{Alvi:2001mx,Poisson:2004cw,Chatziioannou:2012gq} 
\footnote{The tidal Love numbers of slowly spinning BHs are all
zero~\cite{Binnington:2009bb,Landry:2015zfa}.}.
These tidal contributions to the GW phase first appear at $2.5$PN order 
for aligned-spin BBHs, and we keep the leading-order ($2.5$PN) 
and the next-to-leading-order ($3.5$PN)
contributions~\cite{Chatziioannou:2016kem}, 
including the $2.5$PN and $3.5$PN contributions 
due to the energy and angular-momentum fluxes across the BH horizon, 
and the $3.5$PN secular corrections to the binary's binding energy 
and GW luminosity (energy flux emitted to infinity) 
accumulated over the inspiral timescale. 
$\Delta \Psi^{\mathrm {BH-tidal}}_{3.5{\mathrm {PN}}}$ 
is also the function of $v$ and $(\nu,\, \chi_s,\, \chi_a)$, 
which thus adds extra spin-dependent, finite-size contributions 
to $\Psi_{\mathrm {BBH}}(f)$. 
Its explicit expression is derived in Ref.~\cite{Isoyama:2017tbp}; 
note that the tidal heating and torquing for non-spinning BBHs 
start only from $4$PN order, yielding 
$\Delta \Psi^{\mathrm {BH-tidal}}_{3.5 {\mathrm {PN}}}
= 0$~\cite{Chatziioannou:2012gq}.

\subsection{Parameter estimation: Statistical errors}
\label{subsec:statistical}

The parameter errors due to the overall effect of detector noise 
now have a firm statistical foundation 
(see, e.g., Refs.~\cite{Finn:1992wt,Cutler:1994ys}). 
We assume that the GW signal observed in a detector 
(the so-called ``template'') is modeled by the sky-averaged $3.5$PN waveform 
${\tilde h}(f; \boldsymbol{\theta})$ (see Eq.~\eqref{hf} 
with the set of physical parameters $\boldsymbol{\theta}$. 
We also assume that the noise in a detector 
is additive, stationary, Gaussian with zero means, 
and uncorrected with each other when considering 
a multiband network of GW detectors.

We begin with the single detector configuration. 
In the matched filtering analysis, the SNR
(corresponding to the maximum correlation with the optimal filter) 
for the given time-domain signal $h$ is defined by 
\begin{equation}\label{SNR-ave}
\rho_{\mathrm {ave}}
\equiv
\left( h \mid h \right)^{1/2} 
=
\sqrt{\frac{2}{15}}
\frac{{\mathcal {M}}^{5/6}}{D_L} \pi^{-2/3}
\left(
\int_{f_{\mathrm {in}}}^{f_{\mathrm {end}}} 
 \frac{f^{-7/3}}{S_h(f)} \, df 
\right)^{1/2}\,.
\end{equation}
The bracket denotes the inner product weighted 
by the noise power spectrum density $S_h(f)$ 
(asterisk ``$*$'' is used for complex conjugation)~\cite{Finn:1992wt} 
\begin{equation}\label{inner}
\left( a \mid b \right) 
=
2 \int_{f_{\mathrm {in}}}^{f_{\mathrm {end}}} 
\frac{{\tilde a}^* (f) {\tilde b}(f) + {\tilde b}^* (f) {\tilde a}(f)}
{S_h(f)} \, df \,,
\end{equation}
where $[f_{\mathrm {in}},\,f_{\mathrm {end}}]$ is the frequency range 
determined by the detector setup and property of signals; 
see Sec.~\ref{subsec:PSD}.
Note that the SNR of Eq.~\eqref{inner} automatically gives 
the \textit{averaged} SNR over all-sky position 
and binary orientation~\cite{Cutler:1994ys,Dalal:2006qt}, 
because of the ``sky-averaging'' factor of $2/5$ i
n the waveform of Eq.~\eqref{hf}
\footnote{Because all ``sky-averaging'' factors are included 
in the waveform of Eq.~\eqref{hf}, our convention for $S_h(f)$ is 
the standard ``non sky-averaging'': 
$\langle {\tilde n}(f) \,{\tilde n}^*(f') \rangle 
\equiv (1/2) \, \delta (f - f') S_h(f)$, 
where $\delta$ is a delta function, 
${\tilde n}(f)$ is the Fourier component of the noise $n(t)$, 
and angle brackets mean ensemble averaging with respect to 
the noise distribution.}.
Equation~\eqref{SNR-ave} can be recast in terms of $D_{\mathrm {L}}$ 
to describe the observable range for a fixed $\rho_{\mathrm {ave}}$
~\cite{Damour:2000gg,Arun:2004hn}. 
This was used to plot the right panel of FIG.~\ref{fig:PSD-SNR}.

For Gaussian noise and high-SNR sources 
(together with caveats~\cite{Vallisneri:2007ev,Rodriguez:2013mla}),  
the standard Fisher matrix formalism allows us 
to estimate the statistical errors 
$\delta \boldsymbol{\theta} \equiv
\boldsymbol{\theta} - \boldsymbol{\theta}_0$ associated with the measurement, 
where $\boldsymbol{\theta}$ 
and $\boldsymbol{\theta}_0$ are the best-fit parameters 
in the presence of some realization of noise and 
the ``true value'' of the physical parameters, respectively. 
In the high-SNR limit, $\delta \boldsymbol{\theta}$ has 
a Gaussian probability distribution~\cite{Vallisneri:2007ev} 
\begin{equation}\label{p-Gauss}
p(\delta \boldsymbol{\theta}) 
\propto
p^{(0)} (\boldsymbol{\theta}) 
\exp \left( 
-\frac{1}{2} \Gamma_{a b} \delta \theta^a \delta \theta^b
\right)\,,
\end{equation}
where $p^{(0)} (\boldsymbol{\theta})$ 
are the prior probabilities of the physical parameters; 
summation over repeated indices is understood 
(and we do not distinguish upper indices from lower ones).  
Here 
$\Gamma_{a b}
\equiv 
({\partial {\tilde h}}/{\partial \theta_a} 
|
{\partial {\tilde h}}/{\partial \theta_b} )
|_{ {\boldsymbol{\theta}} = {\boldsymbol{\theta}}_0}\,,
$ 
is the Fisher information matrix defined in terms of Eq.~\eqref{inner}, 
and its inverse defines the variance-covariance 
matrix $\Sigma^{a b} \equiv ( \Gamma_{a b} )^{-1}$ 
for the Gaussian distribution of Eq.~\eqref{p-Gauss}. 
Then, the root-mean-square error and cross correlations 
of parameters $\boldsymbol{\theta}$ are given by 
\begin{equation}\label{sta-error}
\sigma_a 
\equiv 
\langle (\delta \theta^a )^2 \rangle^{1/2}
= \sqrt{\Sigma^{aa}}\,, 
\quad
c^{ab} \equiv 
\frac{\langle {\delta \theta^a} {\delta \theta^b} \rangle}{\sigma_a  \sigma_b}
=
\frac{\Sigma^{ab}}{\sqrt{\Sigma^{aa} \Sigma^{bb}}}\,,
\end{equation}
where angle brackets denote an average 
over the Gaussian distribution of Eq.~\eqref{p-Gauss}  
(there is no summation over repeated indices here). 
By definition, each $c^{ab}$ must be restricted 
to the interval $[-1, 1]$; 
when $|c_{ab}| \sim 1$ ($|c_{ab}| \sim 0$), 
it is said that the two parameters $\theta^a$ and $\theta^b$ 
are strongly correlated (almost uncorrelated).

Now we return to a \textit{multiband} network configuration of detectors 
 (e.g., ``B-DECIGO $+$ aLIGO'', and so on). 
Because we assumed that the noise in the different detectors 
is uncorrelated, the total network SNR and 
Fisher information matrix are simply the sum of the individual (averaged) SNRs 
and Fisher information matrix for each detector: 
\begin{equation}\label{Fisher-M}
\rho_{\mathrm {tot}}
\equiv
\sqrt{ (\rho_{\mathrm {ave}}^{\mathrm {I}})^2 
+ (\rho_{\mathrm {ave}}^{\mathrm {II}})^2}\,,
\quad 
{\Gamma}_{ab}^{\mathrm {tot}}
\equiv
{\Gamma}_{ab}^{\mathrm {I}} + {\Gamma}_{ab}^{\mathrm {II}}\,. 
\end{equation}
The total variance-covariance matrix for Eq.~\eqref{p-Gauss} 
is then given by $\Sigma^{a b} \equiv ( \Gamma_{a b}^{\mathrm {tot}} )^{-1}$, 
from which we can estimate the corresponding 
total root-mean-square error and cross correlations of parameters, 
making use of Eq.~\eqref{sta-error}. 
Equations~\eqref{p-Gauss} and~\eqref{Fisher-M} directly show 
the advantage of parameter estimation with the multiband GW network. 
Having a priori knowledge from detector I in a different GW band, 
the parameter estimation with detector II 
could be more precise than a single-band analysis 
using only detector II. 

\subsection{Parameter estimation: Systematic errors}
\label{subsec:systematic}

Next, we collect a few key results from the theory of GW signal analysis 
to measure the systematic \textit{mismodeling} error; 
this arises from the fact that our PN waveform of Eq.~\eqref{hf} used 
in the statistical analysis only approximates 
the true general-relativistic signals.

We focus only on the waveform phasing error 
due to the neglect of the $4$PN non-spinning point-particle term 
in the test-mass limit ($\nu$ = 0), which can over-dominate the error budget 
in measurement of NS tidal effects in the aLIGO
band~\cite{Favata:2013rwa,Yagi:2013baa,Wade:2014vqa}. 
With this setup, we model the ``true'' GW signal 
by the sky-averaged PN waveform 
${\tilde h}_{\mathrm {T}}(f) 
\equiv 
{\cal A} f^{-7/6}  e^{i \Psi_{\mathrm {T}}(f)}$, 
making use of Eq.~\eqref{hf}, and the true GW phase is 
$\Psi_{\mathrm {T}}(f) 
\equiv
\Psi_{\mathrm {BNS/BBH}} 
+
{3} \Delta \Psi^{\mathrm {pp}}_{4{\mathrm {PN}}}/({128 \nu v^5})$,  
where  
\begin{equation}\label{F2-4PN}
\Delta \Psi^{\mathrm {pp}}_{4{\mathrm {PN}}} 
=
\left\{
c_4 \left( \ln(v) -\frac{1}{3} \right) 
+ {\frac {18406}{63}} \ln(v)^2 + O(\nu) \right\} v^8\,. 
\end{equation}
The $\nu$-independent coefficient $|c_4| \sim 3200$ 
is computed in Ref.~\cite{Varma:2013kna} 
built on the results in Refs.~\cite{Fujita:2011zk,Fujita:2012cm};
the calculation of the $\nu$-dependent correction 
to $\Delta \Psi^{\mathrm {pp}}_{4{\mathrm {PN}}}$ 
is a current frontier 
in PN modeling~\cite{Damour:2016abl,Foffa:2016rgu,Marchand:2017pir}.

A standard data-analysis-motivated figure of merit is 
the \textit{match}~\cite{Owen:1995tm,Owen:1998dk} 
that measures the accuracy of the approximate ($3.5$PN) waveform 
${\tilde h} = {\tilde h}_{\mathrm {BNS/BBH}} $ of Eq.~\eqref{hf} 
by comparing to the true waveform ${\tilde h}_{\mathrm {T}}$ 
with identical (true) source parameters $\boldsymbol{\theta}_0$:  
\begin{equation}\label{match}
{\mathrm {match}}
\equiv
\max_{\Delta t_c,\,\Delta \Psi_c } 
{({\hat h}_{\mathrm {T}} \mid {\hat h} )}
=
4\, 
\max_{\Delta t_c} \int_{f_{\mathrm{in}}}^{f_{\mathrm{end}}}
\frac{ {\hat h}_{\mathrm {T}}(f) {\hat h}^{\ast}(f) }
{S_{h}(f)}\, e^{2 \pi i f \Delta t_c} df\,,
\end{equation}
where $\Delta \Psi_c \equiv \Psi_c^{\mathrm {T}} - \Psi_c$,
$\Delta t_c \equiv t_c^{\mathrm {T}} - t_c$, 
and ${\hat h} \equiv {\tilde h} / ({\tilde h} | {\tilde h})^{1/2}$. 
Here, maximizing over the phase $\Delta \Psi_c$ is done
analytically~\cite{Allen:2005fk}. 
Waveform models with low match ($\lesssim 0.97$) are generally considered 
to be not ``faithful'' in the parameter estimation~\cite{Damour:1997ub}.

Still, how much systematic bias on the parameter estimation does 
the high-match ($\gtrsim 0.97$) waveform generate? 
Given a best-fit waveform ${\tilde h} (\boldsymbol{\theta})$ 
to the detector output with the best-fit parameters $\boldsymbol{\theta}$, 
Cutler and Vallisneri showed that 
the systematic estimation errors in the source parameter 
$\Delta \boldsymbol{\theta} 
\equiv
\boldsymbol{\theta} - \boldsymbol{\theta}_0$ 
can be estimated by minimizing the inner product 
$({\hat h}_{\mathrm {T}} (\boldsymbol{\theta}_0) 
- {\hat h} (\boldsymbol{\theta})
|
{\hat h}_{\mathrm {T}} (\boldsymbol{\theta}_0) 
- {\hat h} (\boldsymbol{\theta}))$. 
In the high-SNR regime, it was shown that 
the minimization of this inner product
yields~\cite{Cutler:2007mi} (see also Ref.~\cite{Favata:2013rwa})
\begin{equation}\label{sys-err}
\Delta \theta_{a} 
=
\frac{3}{32} \frac{{\cal A}^{2} (\pi m)^{5/3}}{\nu}
\Gamma_{ab}^{-1} 
\int_{f_{\mathrm {in}}}^{f_{\mathrm {end}}} 
\frac{f^{-2/3}}{S_h(f)} 
\Delta \Psi^{\mathrm {pp}}_{4{\mathrm {PN}}} \,
\partial_b  \Psi \, df \,,
\end{equation}
where we use the sky-averaged PN waveform of Eq.~\eqref{hf} 
with the assumption $\Delta \theta^{a} \partial_a \Psi \lesssim 1$.
In contrast to the statistical errors 
$\boldsymbol{\sigma} \sim {\mathrm {SNR}}^{-1}$, 
it is important to recognize that 
the systematic error $\Delta \theta_{a}$ is essentially 
\textit{independent of SNR} because both the SNR of Eq.~\eqref{SNR-ave} 
and the Fisher information matrix scale as 
${\mathrm {SNR}}^2 \sim {\cal A}^2 \sim  \Gamma_{ab}$, 
while $\Delta \theta_{a} \sim {\cal A}^0$.

\subsection{Noise sensitivity of B-DECIGO and other GW detectors}
\label{subsec:PSD}

The computation of the statistical and systematic parameter estimation errors 
will require as an input the model of noise power spectrum density $S_h(f)$ 
corresponding to each GW detector. 
For B-DECIGO, we use the expected $S_h(f)$ proposed by 
Nakamura et al. \cite{Nakamura:2016hna} 
(see also Ref.~\cite{Yagi:2011wg} for the (original) DECIGO configuration)
\begin{equation}\label{S-BD}
S_{h}(f) \equiv S_0 
\left( 1.0 + 1.584 \times 10^{-2}\, y^{-4} 
+ 1.584 \times 10^{-3}\, y^2 \right)\,
\end{equation}
with $y \equiv f / (1.0 \,{\mathrm{Hz}})$,  
$S_0 \equiv 4.040 \times 10^{-46} \,{\mathrm{Hz^{-1}}}$, 
and our default frequency range 
$[f_{\mathrm {low}},\, f_{\mathrm {up}}]
=
[0.01,\, 1.0 \times 10^2]\,{\text{Hz}}$. 
Note that $S_0$ may include the geometrical factor 
$3/4 = \sin^2 (\pi/3)$ due to the $60^{\circ}$ opening angle 
of the constellation.

For other GW bands, we consider aLIGO~\cite{Ajith:2011ec}  
and a 3rd generation GW detector, 
the Einstein Telescope (ET)~\cite{Sathyaprakash:2009xs}, 
in the hecto-Hz band as well as LISA~\cite{Babak:2017tow} in the milli-Hz band 
(see also Ref.~\cite{Cornish:2018dyw}). 
Their analytic fits to $S_h(f)$ can be found in the cited references,  
and we choose their default frequency interval as 
$[f_{\mathrm {low}},\, f_{\mathrm {up}}]
=
\{[10.0,\, 2.0 \times 10^3],\,
[2.0,\, 2.0 \times 10^3],\, [1.0 \times 10^{-4},\, 1.0] 
\}\,{\text{Hz}}$, respectively 
\footnote{The factor $20/3 = 5 / \sin^2 (\pi/3)$ in Eq.~(2.1) 
of Ref.~\cite{Babak:2017tow} is a standard conversion factor 
between aLIGO and LISA configurations, 
which is (the inverse of) the products of $1/5$ from an average over 
the pattern functions and $3/4 = \sin^2 (\pi/3)$ from the angle 
between detector arms being $60^{\circ}$~\cite{Barack:2004wc}. 
Recall that our convention for $S_h(f)$ is 
``non-sky-averaging'' because the amplitude parameter ${\cal A}$ 
in Eq.~\eqref{hf} has already accounted for the pattern-average factor $1/5$. 
To avoid counting this factor twice, 
we import $S_h(f)$ from Ref.~\cite{Babak:2017tow} after replacing 
the factor $20/3$ with $4/3$.}. 
Their amplitude spectral densities $\sqrt{S_h(f)}\,{\mathrm {Hz}^{-1/2}}$ 
within $[f_{\mathrm {low}},\, f_{\mathrm {up}}]$ are plotted 
in the left panel of FIG.~\ref{fig:PSD-SNR}; 
note that we shall not consider the galactic confusion noise component 
in the milli-Hz band~\cite{Klein:2015hvg,Audley:2017drz}.

We assume a $T_{\mathrm {obs}} = 4 \, {\mathrm{yr}}$ observation period 
(except in FIG.~\ref{fig:BD-band}, as specified),  
which echoes the mission lifetime requirement of LISA~\cite{Audley:2017drz}. 
All waveforms have cutoff frequencies at
$
{f_{\mathrm {in}}} = 
\max
\{
1.65 \times 10^{-2}
({{\cal M}}/{30 M_{\odot}})^{-5/8} 
({T_{\mathrm {obs}}}/{4 \,{\mathrm{yr}}})^{-3/8},\, 
f_{\mathrm {low}}
\}
$~\cite{Berti:2004bd} and
${f_{\mathrm {end}}} = 
\min \{ f_{\mathrm {ISCO}},\, f_{\mathrm {up}}\}$,
where the ISCO frequency of the Schwarzschild metric 
$f_{\mathrm {ISCO}}$ of Eq.~\eqref{f-ISCO} is determined 
by the redshifted total mass $m$ 
\footnote{The abrupt cutoff of the waveform at $f_{\mathrm {ISCO}}$ 
could artificially improve the parameter estimation 
if the waveform has sufficient noise-weighted power 
at the ISCO frequency~\cite{Mandel:2014tca}. 
For simplicity, we ignore this systematic bias.}.

\subsection{Binary parameters}
\label{subsec:system}

In our simplified version of binary problems, 
the sky-averaged PN waveforms ${\tilde h}_{\mathrm {BNS}}$ 
and ${\tilde h}_{\mathrm {BBH}}$ in Eq.~\eqref{hf} 
depend on $11$- and $7$-dimensional parameters, respectively: 
\begin{align}\label{theta}
\boldsymbol{\theta}_{\mathrm {BNS}} 
&= (\ln{\cal A},\, f_0\,t_c,\, \Psi_c,\, \ln m,\, \nu,\, \chi_s,\, \chi_a,\, 
{\tilde {\mathcal{Q}}},\,\delta {\tilde {\mathcal {Q}}},\,
{\tilde {\Lambda}},\, \delta {\tilde {\Lambda}})\,, \cr 
\boldsymbol{\theta}_{\mathrm {BBH}} 
&= (\ln{\cal A},\, f_0\,t_c,\, \Psi_c,\, \ln m,\, \nu,\, \chi_s,\, \chi_a)\,.
\end{align}
Here, we absorb all amplitude information 
into the single parameter ${\cal A}$, 
and set $f_0 = 1.65 \,{\mathrm {Hz}}$, 
at which $S_h(f)$ of B-DECIGO is minimum.  
However, it is known that the PN waveform of Eq.~\eqref{hf} yields 
the block diagonal form of the Fisher information matrix 
$\Gamma_{\ln {\cal A} \, a} 
= \delta_{\ln {\cal A}\, a} \rho_{\mathrm {ave}}^2$, 
which means that $\ln {\cal A}$ is entirely uncorrelated 
with the other parameters~\cite{Poisson:1995ef}. 
For this reason, we shall remove $\ln{\cal A}$ 
from our list of independent parameters 
and only consider the other $6$-dimensional parameters, 
assuming that they are unconstrained (with ``flat'' priors 
$p^{(0)}(\boldsymbol{\theta}) \sim {\mathrm {const.}}$ in Eq.~\eqref{p-Gauss})
\footnote{It should be borne in mind that in general a different 
prior assumption leads to different conclusions on the parameter estimation 
(see, e.g., Refs.~\cite{Poisson:1995ef,Vitale:2017cfs}).
From this point of view, it would be more physical to take into account priors 
for the fact that $\chi_{1,2}$ and $\nu$ are restricted 
to the interval $[-1, 1]$ and $(0, 1/4]$, respectively. 
Analysis that assumes a certain prior 
for ${\tilde \Lambda}$ and $\delta {\tilde \Lambda}$ 
would also improve their estimation~\cite{Agathos:2015uaa,TheLIGOScientific:2017qsa}.}.

For BNSs, we have parameterized NS quadrupole-moment and tidal effects 
in terms of ``combined'' dimensionless parameters 
$({\tilde {\mathcal{Q}}},\,\delta {\tilde {\mathcal{Q}}})$
in Eq.~\eqref{def-tQ}
and $({\tilde \Lambda},\, \delta {\tilde \Lambda})$ in Eq.~\eqref{def-tL}, 
instead of $\kappa_{1,2}$ and ${\Lambda}_{1,2}$, respectively,  
to improve their measurement
precision~\cite{TheLIGOScientific:2017qsa,Abbott:2018wiz}. 
Furthermore, we shall exclude the parameters 
$(\delta {\tilde {\mathcal{Q}}}, \delta {\tilde \Lambda})$ 
from our estimation. 
The rationale for our choice is that 
$(\delta {\tilde {\mathcal{Q}}}, \delta {\tilde \Lambda})$ 
only show up as the next-to-leading-order corrections in the GW phase 
(recall Eqs.~\eqref{NS-QM} and~\eqref{NS-tidal}), 
and their contributions can be very small 
\footnote{
For the GW170817-like equal-spin BNS (System A) described below, 
we have $\delta {\tilde {\mathcal{Q}}} / {\tilde {\mathcal{Q}}} \sim 0.07$ 
and $\delta {\tilde {\Lambda}} / {\tilde {\Lambda}} \sim 0.08$, 
and $(\delta {\tilde {\mathcal{Q}}}, \delta {\tilde \Lambda})$ 
in the GW phase are further suppressed 
by the factor $\sqrt{1 - 4 \nu} \sim 0.04$; 
they are essentially negligible in our analysis.}.

In this work er consider the following three binaries for B-DECIGO; 
the parameter of System C is motivated by LISA's ``threshold'' system 
(with a source-frame total mass of $m^S \sim 3000 M_{\odot}$ 
and mass ratio of $m_1 / m_2 = 0.2$) 
that defines LISA's observational requirement~\cite{Audley:2017drz}. 
\begin{itemize}
\item  System A: 
GW170817-like BNS with individual masses of 
$(m_1 = 1.3 M_{\odot},\, m_2 = 1.4 M_{\odot})$
and dimensionless spin magnitudes of $\chi_1 = \chi_2 = 0.05$ 
as well as a source-frame dimensionless quadrupole parameter of 
${\tilde {\mathcal {Q}}}^S = 1.75 \times 10^{-2}$
and tidal deformability of ${\tilde \Lambda}^S = 7.03 \times 10^2$, 
located at $40\,{\mathrm {Mpc}}$ ($z \sim 0.009$).  
\item System B: 
GW150914-like BBH with individual masses of 
$(m_1 = 30 M_{\odot},\, m_2 = 40 M_{\odot})$ 
and dimensionless spin magnitudes of 
$(\chi_1 = 0.9,\,\chi_2 = 0.7)$, 
located at $400\,{\mathrm {Mpc}}$ ($z \sim 0.09$). 
\item System C: 
LISA's ``threshold'' BBH with individual masses of 
$(m_1 = 1000 M_{\odot},\, m_2 = 5000 M_{\odot})$ 
and dimensionless spin magnitudes of 
$(\chi_1 = 0.9,\,\chi_2 = 0.7)$, 
located at $6.8 \,{\mathrm {Gpc}}$ ($z \sim 1.0$). 
\end{itemize}
The coalescence time and phase are $t_c = 0.0 = {\Psi}_c$ for each system.
When we measure them using multiband network detectors, 
systems A, B, and C are also visible 
in the aLIGO band, aLIGO and LISA band, and LISA band, respectively 
(recall FIG.~\ref{fig:PSD-SNR}).

\section{Results of parameter estimation}
\label{sec:result}

In what follows we summarize our parameter estimation results 
for each system.
We stress that the methodology of our analysis is extremely simplified,
e.g., with a flat prior and by using only the inspiral-only PN waveform; 
indeed, our aLIGO estimation errors for System A (``GW170817'')
and System B (``GW150914'') obviously contradict those measured 
by aLIGO and advanced Virgo~\cite{Abbott:2016blz, TheLIGOScientific:2017qsa}.  
Thus, the estimation errors that we quote below should be 
\textit{only indicative and tentative}.
Our results have to be followed up 
by using more accurate inspiral-merger-ringdown waveforms 
with additional known effects 
(e.g., NS spins, spin-induced precession, etc.) 
and the orbital motion of B-DECIGO, 
a more rigorous Baysian-posterior-based parameter estimation method,  
and extended to a more exhaustive study of parameter spaces in future. 

However, we are confident that the overall trend of our results 
(e.g., the order of magnitude of errors) should be robust, 
and provides a realistic idea of what can be measured with B-DECIGO 
as well as the multiband network including it.

\subsection{Statistical errors: Fixed SNR}
\label{subsec:sta_err}

The statistical parameter estimation errors scale as 
${\boldsymbol {\sigma}} \propto 
{\mathrm {SNR}}^{-1} (1
+ O({\mathrm {SNR^{-1}}}))$~\cite{Vallisneri:2007ev,Rodriguez:2013mla}, 
and their achievable precision depends on \textit{both} 
the SNR and the bandwidth, over which the SNR is accumulated. 
For the former aspect, B-DECIGO measurement has already 
shown an advantage in FIG.~\ref{fig:PSD-SNR}, where we find that 
System A, B, and C are all high-SNR ($\sim 10^2$) sources. 
Furthermore, the multiband measurement does better than B-DECIGO alone 
because the systems considered are always much louder 
in the network SNR~\eqref{Fisher-M}.

Here, we look at the latter aspects of the statistical errors, 
i.e., their improvement arising from the broader bandwidth. 
To best quantify this, we introduce the \textit{normalized statistical errors} 
$\delta {\boldsymbol {\hat \theta}}$ normalized to a fixed SNR $\rho$, 
\begin{equation}\label{hat-theta}
\delta {\boldsymbol {\hat \theta}}
\equiv 
\rho\,{\boldsymbol {\sigma}} \,,
\end{equation}
and display the statistical errors in terms of 
$\delta {\boldsymbol {\hat \theta}}$; 
the overall statistical errors are simply recovered from 
${\boldsymbol {\sigma}} 
= 
\delta {\boldsymbol {\hat \theta}} / \rho$.

\begin{figure}[tbp]
\begin{tabular}{cc}  
\begin{minipage}[t]{.47\hsize}
  \centering
  \includegraphics[width=\columnwidth, bb = 24 33 706 522]{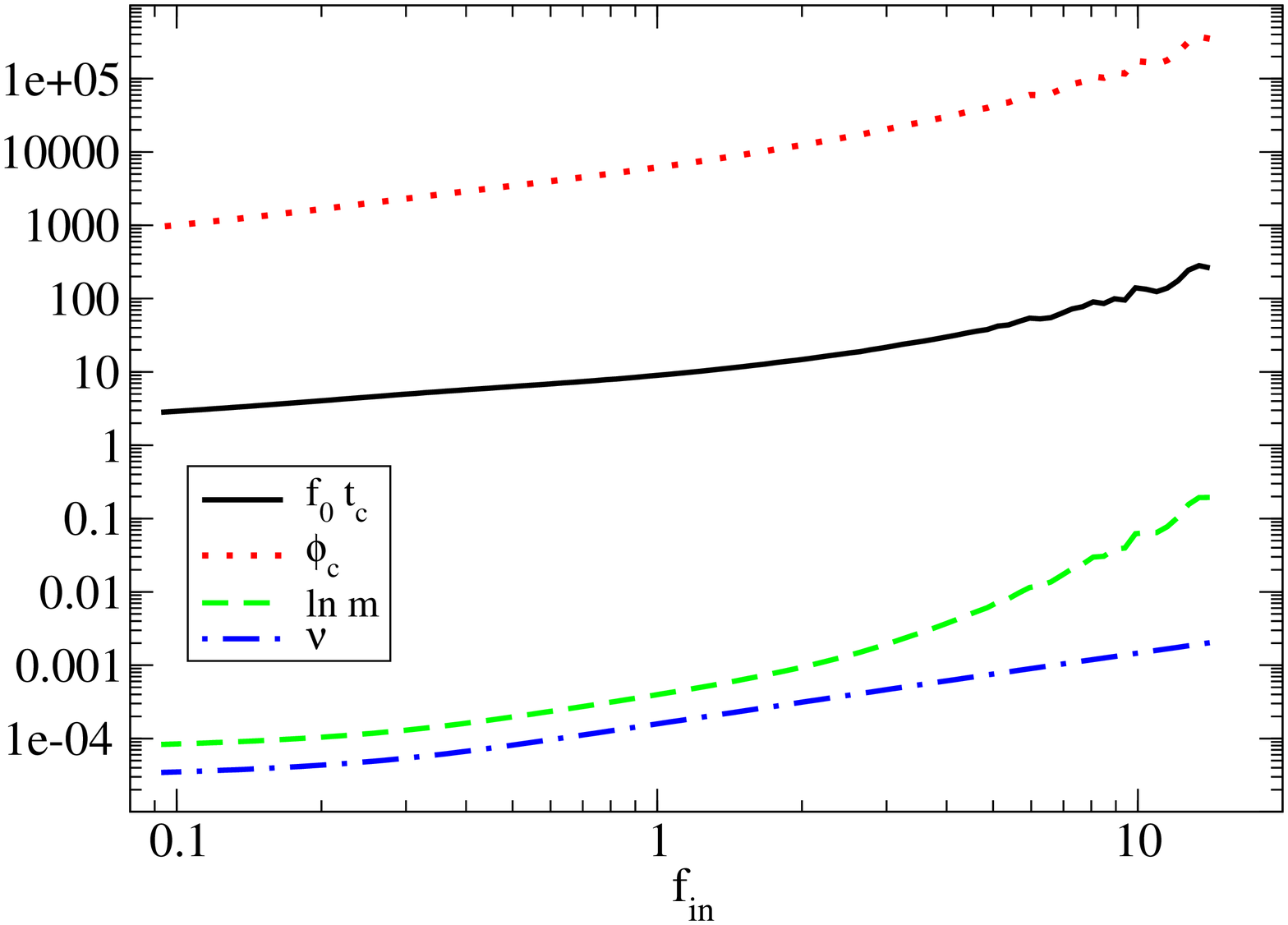}
\end{minipage}
\qquad 
\begin{minipage}[t]{.47\hsize}
  \centering
  \includegraphics[width=\columnwidth, bb = 24 33 706 522]{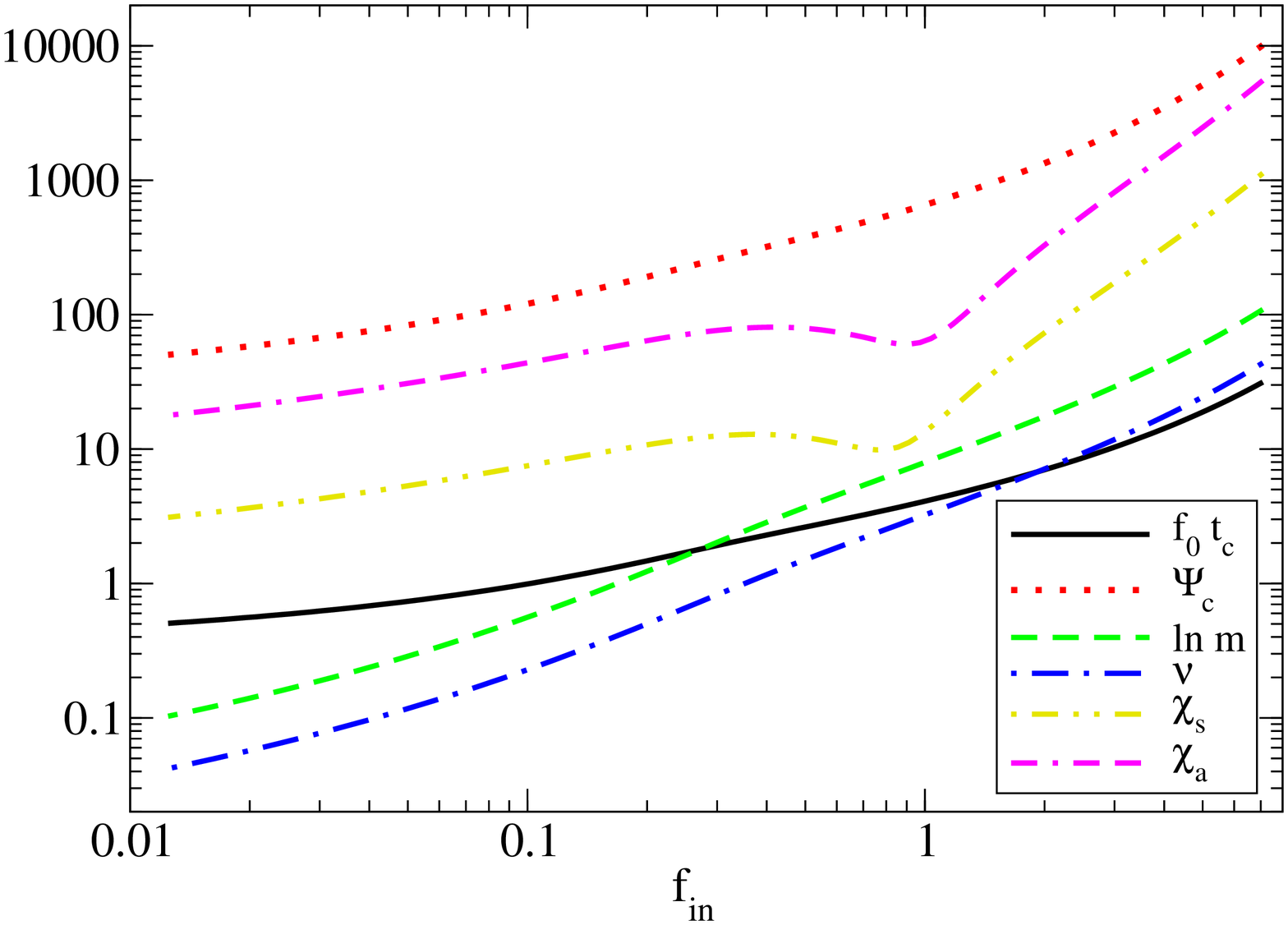}
  \end{minipage}
\end{tabular}
\caption{
B-DECIGO statistical errors 
$\delta {\boldsymbol {\hat \theta}}_{\mathrm {BD}}$ 
normalized to its SNR $\rho^{\mathrm {BD}}_{\mathrm {ave}}$ 
for System A 
(GW170817-like BNS with $\rho^{\mathrm {BD}}_{\mathrm {ave}} = 1.67 \times 10^{2}$: 
\textit{left})  
and System B 
(GW150914-like BBH with $\rho^{\mathrm {BD}}_{\mathrm {ave}} = 2.51 \times 10^{2}$: 
\textit{right}), 
as a function of the lower cut-off frequency ${f_{\mathrm {in}}}$; 
we set $f_0 = 1.65 \,{\mathrm {Hz}}$ 
at which B-DECIGO $S_h(f)$ is minimum. 
The errors $\delta \chi_{s,a}$ 
and $(\delta {\hat {\mathcal Q}},\,\delta {\hat {\Lambda}})$ 
for System A are not plotted as B-DECIGO cannot constrain them. 
The range of ${f_{\mathrm {in}}}$ corresponds to a 
$[7\,{\mathrm {min}}, 10\,{\mathrm{yr}}]$ observation (System A) 
and 
$[0.2\,{\mathrm {min}}, 10\,{\mathrm{yr}}]$ observation (System B) 
before reaching the ISCO frequency ${f_{\mathrm {ISCO}}}$; 
we have 
$f_{\mathrm {in}} \sim 1.2 \times 10^{-1} \,{\mathrm {Hz}}$ (System A) 
and 
$ \sim 1.7 \times 10^{-2} \,{\mathrm {Hz}}$ (System B)
for a $4 \, {\mathrm{yr}}$ observation. } 
\label{fig:BD-band}
\end{figure}

We start by looking at System A (GW170817-like BNS). 
In the left panel of FIG.~\ref{fig:BD-band}, 
we show the B-DECIGO statistical errors 
$\delta {\boldsymbol {\hat \theta}}_{\mathrm {BD}}$ normalized to its SNR 
$\rho^{\mathrm {BD}}_{\mathrm {ave}}$ (see Table~\ref{table:sta-errA}) 
as a function of the lower cutoff frequency ${f_{\mathrm {in}}}$. 
Here we leave the errors $\delta \chi_{s,a}$ 
and $(\delta {\hat {\mathcal Q}},\,\delta {\hat {\Lambda}})$ 
out of the plot because B-DECIGO is not able 
to discern the parameters associated with the waveforms; 
they are significant only in the higher aLIGO band, as anticipated. 
We see that $\delta {\boldsymbol {\hat \theta}}_{\mathrm {BD}}$ 
becomes significantly smaller 
as the signal contributes to broader bandwidth 
\footnote{We find that $\delta {\hat \theta}_{\mathrm {BD}}^a$ 
in FIG.~\ref{fig:BD-band} can produce ``bumps'' 
when the corresponding covariance matrix $c^{ab}$ of Eq.~\eqref{sta-error} 
changes their signs, crossing zero. 
When $c^{ab} = 0$, ${\theta}^{a}$ and ${\theta}^{b}$ become 
entirely uncorrelated with each other, and this suddenly 
``improves'' (or ``worsens'') the parameter estimation.  
In our investigation, this should be another systematic bias 
due to PN waveforms because the varying PN order of the GW phase 
$\Psi_{\mathrm {BNS/BBH}}$ in Eqs.~\eqref{F2-BNS} and~\eqref{F2-BBH} 
shifts the location of these bumps (for fixed binary parameters). 
We postpone a more detailed study of this issue to the future, 
but their systematic nature should lend caution 
to future parameter-estimation studies of binary inspirals 
based on the Fisher information matrix and PN waveforms.}.
For a $4 \, {\mathrm{yr}}$ observation before the coalescence, e.g., 
${\delta {\hat \nu}} / \nu$ is below $0.05\%$ mark, and 
the overall statistical error in $t_c$ is 
$\sigma_{t_c} \lesssim 5.0\,{\mathrm s}$ 
when ${f_{\mathrm {end}}} = 1.0\,{\mathrm Hz}$ 
(corresponding to $\sim 6$ d before reaching at $f_{\mathrm {ISCO}}$).
Hence, B-DECIGO is able to alert aLIGO and electromagnetic observatories 
about the time of merger well in advance.

Table~\ref{table:sta-errA} reports $\delta {\boldsymbol {\hat \theta}}$ 
for System A measured by two different multiband GW networks, 
``B-DECIGO  $+$ aLIGO'' (the first column) 
and ``B-DECIGO $+$  ET'' (the second column); 
they are now normalized to the corresponding 
total network SNR $\rho_{\mathrm {tot}}$.
We see that ${\delta {\hat m}}$ and ${\delta {\hat \nu}}$ in the B-DECIGO 
measurements can be far better than those in aLIGO (ET) measurements 
by $\sim 2$ orders of magnitude. 
For the multiband measurement with B-DECIGO $+$ aLIGO (ET), 
the improvement in ${\delta {\hat m}}$ and ${\delta {\hat \nu}}$ 
is only incremental as B-DECIGO already measures them quite precisely 
during the early inspiral. 
When it comes to the NS (symmetric) spins $\chi_s$ and 
matter imprints like quadrupole and tidal parameters 
$({\tilde {\mathcal Q}},\,{\tilde \Lambda})$, however, 
the benefit from having B-DECIGO information becomes drastic.
The multiband analysis can reduce the error 
in the tidal parameter $\delta {\hat \Lambda} / {\tilde \Lambda}$
by an order of magnitude in aLIGO or ET alone. 
Furthermore, it is interesting to observe that 
there is the potential to extract 
the spin and spin-induced quadrupole parameters 
$(\chi_s,\,{\tilde {\mathcal {Q}}})$ from the GW waveforms 
making use of B-DECIGO $+$ aLIGO (ET) measurements. 
While the errors $(\delta {\hat \chi_s} / \chi_s 
,\,\delta {\hat {\mathcal {Q}}} / {\tilde {\mathcal {Q}}})$ 
determined by only B-DECIGO or aLIGO(ET) 
are too large to yield any constraint on them, 
the multiband measurement significantly reduces these errors 
by two orders of magnitude. 
Indeed, these results benefit from the mass ratio 
being very precisely measured by B-DECIGO. 
Our estimation suggests that in the future B-DECIGO $+$ 
aLIGO (ET) multiband era 
it will allow us to measure the NS equation of state much better 
relying on \textit{both} spin and tidal effects.  

\begin{table}[tb]
\begingroup
\renewcommand{\arraystretch}{1.2} 
\scalebox{0.63}[0.63]{
\begin{tabular}{lc|ccccccc}
\hline\hline
Detector  \quad & \quad SNR & 
\quad $\delta {\hat t}_c$ & \quad $\delta {\hat \Psi}_c$ & 
\quad ${\delta {\hat m}} / m$ & \quad ${\delta {\hat \nu}} / \nu$ & 
\quad $\delta {\hat \chi_s} / \chi_s$  & 
\quad $\delta {\hat {\mathcal {Q}}} / {\tilde {\mathcal {Q}}}$  & 
\quad $\delta {\hat \Lambda} / {\tilde \Lambda}$ \\
\hline
\multicolumn{9}{l}{B-DECIGO (BD) + aLIGO}\\
\hline
BD  \quad & \quad ($1.67 \times 10^{2}$) &
\quad $1.98$ & \quad $1.19 \times 10^{3}$ & 
\quad $1.03 \times 10^{-4}$ & \quad $1.71 \times 10^{-4}$ & 
\quad $\cdots$ & \quad $\cdots$ & \quad $\cdots$ \\ 
aLIGO  \quad & \quad ($3.49 \times 10^{1}$) &
\quad $1.16$ & \quad $4.05 \times 10^{3}$ & 
\quad $6.19 \times 10^{-2}$ & \quad $9.93 \times 10^{-2}$ & 
\quad $7.87 \times 10^{3}$ & 
\quad $2.44 \times 10^{4}$ & \quad $1.35 \times 10^{2}$ \\ 
BD + aLIGO  \quad & \quad $1.71 \times 10^{2}$ &
\quad $5.11 \times 10^{-2}$ & \quad $1.94$ & 
\quad $1.03 \times 10^{-4}$ & \quad $1.71 \times 10^{-4}$ & 
\quad $1.81 \times 10^{2}$ & 
\quad $1.73 \times 10^{2}$ & \quad $2.13 \times 10^{1}$ \\ 
\hline
\multicolumn{9}{l}{B-DECIGO (BD) + ET}\\
\hline
BD  \quad & \quad ($1.67 \times 10^{2}$) &
\quad $5.93$ & \quad $3.56 \times 10^{3}$ & 
\quad $3.07 \times 10^{-4}$ & \quad $5.12 \times 10^{-4}$ & 
\quad $\cdots$ & \quad $\cdots$ & \quad $\cdots$ \\ 
ET \quad & \quad ($4.83 \times 10^{2}$) &
\quad $1.31 \times 10^{-1}$ & \quad $4.01 \times 10^{2}$ & 
\quad $1.18 \times 10^{-2}$ & \quad $1.97 \times 10^{-2}$ & 
\quad $8.45 \times 10^{2}$ & 
\quad $1.99 \times 10^{3}$ & \quad $2.17 \times 10^{1}$ \\ 
BD + ET  \quad & \quad $5.11 \times 10^{2}$ &
\quad $2.65 \times 10^{-2}$ & \quad $6.39 \times 10^{1}$ & 
\quad $3.07 \times 10^{-4}$ & \quad $5.12 \times 10^{-4}$ & 
\quad $1.22 \times 10^{2}$ & 
\quad $1.50 \times 10^{2}$ & \quad $8.85$ \\ 
\hline\hline
\end{tabular}
}
\endgroup
\caption{Statistical parameter errors 
$\delta {\boldsymbol {\hat \theta}}$ normalized to the total multiband SNR 
$\rho_{\mathrm {tot}}$ for System A (GW170817-like BNS). 
The first and second columns give $\delta {\boldsymbol {\hat \theta}}$ 
for the two different multiband networks B-DCEIGO $+$ aLIGO  
and B-DECIGO $+$ ET, respectively 
(thus, using different total SNRs $\rho_{\mathrm {tot}}$ 
for $\delta {\boldsymbol {\hat \theta}}$).
For each of the two columns, the first and second lines 
show $\delta {\boldsymbol {\hat \theta}}$ 
using only B-DECIGO and aLIGO (or ET), respectively, 
while the third line displays those obtained by multiband measurement. 
They are evaluated at the true values $\nu = 0.249657$, 
$\chi_s = 0.05$, and $\chi_a = 0.0$. 
The blank cells indicate at least 
$| \delta {\boldsymbol {\hat \theta}} / \boldsymbol{\theta}_{\mathrm {BNS}} | 
> 10^3$, and the error $\delta {\hat \chi_a}$ are not displayed here 
because none of the detectors or networks listed here can measure it 
(except B-DECIGO + ET network, which yields 
$\delta {\hat \chi_a} \sim 1.37 \times 10^{2}$).}
\label{table:sta-errA}
\end{table}
%

We next turn to examine System B (GW150914-like BBH). 
In the right panel of FIG.~\ref{fig:BD-band} 
we plot $\delta {\boldsymbol {\hat \theta}}_{\mathrm {BD}}$ 
normalized to the B-DECIGO SNR $\rho^{\mathrm {BD}}_{\mathrm {ave}}$
(see Table~\ref{table:sta-errB}) as a function of ${f_{\mathrm {in}}}$. 
The overall trend of $\delta {\boldsymbol {\hat \theta}}_{\mathrm {BBH}}$ 
is similar as that of the left panel (System A), 
but, strikingly, \textit{B-DECIGO is able to measure 
the individual BH spins ${\chi_{1,2}}$}; 
after converting $\delta {\boldsymbol {\hat \chi}}_{s,a}^{\mathrm {BD}}$ 
to the overall statistical errors of BH spins $\sigma_{\chi_{1,2}}$ 
for $\rho^{\mathrm {BD}}_{\mathrm {ave}}$) 
with the standard variance-propagation formula 
\footnote{Suppose random variables are collected in the random vector 
${\boldsymbol {y}}$, together with a non-linear function 
${\boldsymbol {y}} = {\boldsymbol  f}({\boldsymbol {x}})$ 
of Gaussian random variables ${\boldsymbol {x}}$. 
In a first-order approximation of a Taylor series expansion 
of ${\boldsymbol {y}} 
=
{\boldsymbol f}({\boldsymbol {x}_0}) 
+ 
{\boldsymbol J} d {\boldsymbol {x}} \dots
$ 
around a given vector ${\boldsymbol {x}_{0}}$ 
with Jacobian matrix $J_{ab} = (\partial f_a(x)) / {\partial x_b} 
|_{{x} = {x_0}}$, 
the variance-covariance matrix ${\boldsymbol {\Sigma}}_{yy}$
for ${\boldsymbol {y}}$ is given by 
${\boldsymbol {\Sigma}}_{yy} 
= 
{\boldsymbol{J}} {\boldsymbol{\Sigma}}_{xx} {\boldsymbol{J}}^T$
where ${\boldsymbol{\Sigma}}_{xx}$ is the variance-covariance matrix 
for ${\boldsymbol {x}}$.},
we find that $\chi_{1,2}$ can be measured better than $10\%$ 
(for a $4 \, {\mathrm{yr}}$ observation).  
This result seems counterintuitive because the spin effects 
in the GW phase first enter through the $1.5$ PN spin-orbit couplings~\cite{Blanchet:2013haa}, expected to emerge in the high frequency aLIGO band. 
However, recall that the spin-dependent point-particle 
contribution $\Delta \Psi^{\mathrm {pp}}_{3.5{\mathrm {PN}}}$ 
in Eqs.~\eqref{F2-BNS} and~\eqref{F2-BBH} are proportional to 
the inverse of $v^{-5}$. Because of that, 
the $1.5$PN spin-orbit term could actually be 
more pronounced in the lower B-DECIGO band, 
introducing greater variety in the waveform.
This is indeed what the results show. 
We remark that the spin-dependent tidal corrections  
$\Delta \Psi^{\mathrm {BH-tidal}}_{3.5{\mathrm {PN}}}$ 
play an important role for B-DECIGO measurement from this point of view; 
its neglect adds $\sim 2\%$ relative uncertainties 
when measuring ${\hat \chi_{s,a}}$.

Table~\ref{table:sta-errB} reports $\delta {\boldsymbol {\hat \theta}}$ 
for System B measured by three different multiband GW networks, 
``B-DECIGO  $+$ aLIGO'' (the first column), 
``B-DECIGO $+$  ET'' (the second column), 
and ``LISA $+$  aLIGO'' (the third column), 
each normalized to the total network SNR $\rho_{\mathrm {tot}}$. 
We see that the B-DECIGO measurement of 
${\boldsymbol {\hat \theta}}_{\mathrm {BBH}}$ 
can be all improved tremendously, 
by $3$ to $4$ ($1$ to $2$) orders of magnitude, 
compared to the aLIGO (ET) measurement
\footnote{Recall that they are overestimated 
due to our simplified inspiral-only analysis in the aLIGO band. 
If we instead compare our B-DECIGO results 
with the measurement uncertainties of
GW150914~\cite{TheLIGOScientific:2016wfe}, 
the improvement against the aLIGO measurement may be 
by $\sim 2$ orders of magnitude.}.
Although ${\boldsymbol {\hat \theta}}_{\mathrm {BBH}}$ 
is largely constrained in the B-DECIGO band, 
there is further several factors of improvement 
for joint B-DECIGO $+$ aLIGO (ET) analysis. 
Also, the broader conclusions that we can draw from 
``LISA $+$  aLIGO'' measurement agree with those 
by Sasana~\cite{Sesana:2016ljz} and Vitale~\cite{Vitale:2016rfr}, 
which focused on the similar GW150914-like system; 
LISA can predict the coalescence time within less than $30{\mathrm {s}}$, 
and prior information from LISA can reduce the uncertainties
of individual BH spins in the aLIGO measurement
(we shall not, however, attempt a precise comparison with them 
because of the significant difference in methodology). 
In principle, we could even consider full multiband measurement 
with a ``LISA $+$ B-DECIGO $+$ aLIGO (ET)'' network. 
However, it makes virtually no difference compared 
to the B-DECIGO $+$ aLIGO (ET) network. 
B-DECIGO already measures ${\boldsymbol {\hat \theta}}_{\mathrm {BBH}}$ 
very precisely, and the LISA contribution to $\rho_{\mathrm {tot}}$ 
is very small.

\begin{table}[tb]
\begingroup
\renewcommand{\arraystretch}{1.2} 
\scalebox{0.7}[0.7]{
\begin{tabular}{lc|cccccc}
\hline\hline
Detector  \quad & \quad SNR & 
\quad $\delta {\hat t}_c$ & \quad $\delta {\hat \Psi}_c$ & 
\quad ${\delta {\hat m}} / m$ & \quad ${\delta {\hat \nu}} / \nu$ & 
\quad $\delta {\hat \chi_s} / \chi_s$  & 
\quad $\delta {\hat \chi_a} / \chi_a$ \\
\hline
\multicolumn{8}{l}{B-DECIGO (BD) + aLIGO}\\
\hline
BD  \quad & \quad $(2.51 \times 10^{2})$ &
\quad $3.27 \times 10^{-1}$ & \quad $5.52 \times 10^{1}$ & 
\quad $1.24 \times 10^{-1}$ & \quad $2.07 \times 10^{-1}$ & 
\quad $4.31$ & \quad $1.98 \times 10^{2}$ \\ 
a-LIGO  \quad & \quad $(3.70 \times 10^{1})$ &
\quad $1.94 \times 10^{2}$ & \quad $1.48 \times 10^{5}$ & 
\quad $1.25 \times 10^{3}$ & \quad $2.01 \times 10^{3}$ & 
\quad $2.31 \times 10^{4}$ & \quad $9.15 \times 10^{5}$ \\ 
BD + aLIGO  \quad & \quad $2.54 \times 10^{2}$ &
\quad $1.39 \times 10^{-1}$ & \quad $3.48 \times 10^{1}$ & 
\quad $9.86 \times 10^{-2}$ & \quad $1.61 \times 10^{-1}$ & 
\quad $2.89$ & \quad $1.33 \times 10^{2}$ \\ 
\hline
\multicolumn{8}{l}{B-DECIGO (BD) + ET}\\
\hline
BD  \quad & \quad $(2.51 \times 10^{2})$ &
\quad $6.70 \times 10^{-1}$ & \quad $1.13 \times 10^{2}$ & 
\quad $2.53 \times 10^{-1}$ & \quad $4.22 \times 10^{-1}$ & 
\quad $8.78$ & \quad $4.03 \times 10^{2}$ \\ 
ET \quad & \quad $(4.53 \times 10^{2})$ &
\quad $3.97$ & \quad $1.83 \times 10^{3}$ & 
\quad $2.21 \times 10^{1}$ & \quad $3.64 \times 10^{1}$ & 
\quad $2.46 \times 10^{2}$ & \quad $9.58 \times 10^{3}$ \\ 
BD + ET  \quad & \quad $5.18 \times 10^{2}$ &
\quad $8.17 \times 10^{-2}$ & \quad $1.90 \times 10^{1}$ & 
\quad $1.23 \times 10^{-1}$ & \quad $2.05 \times 10^{-1}$ & 
\quad $2.10$ & \quad $9.68 \times 10^{1}$ \\ 
\hline
\multicolumn{8}{l}{LISA + aLIGO}\\
\hline
LISA  \quad & \quad $(5.16)$ &
\quad $1.19 \times 10^{3}$ & \quad $2.23 \times 10^{4}$ & 
\quad $5.64$ & \quad $9.39$ & 
\quad $1.00 \times 10^{3}$ & \quad $4.63 \times 10^{4}$ \\ 
a-LIGO  \quad & \quad $(3.70 \times 10^{1})$ &
\quad $2.86 \times 10^{1}$ & \quad $2.18 \times 10^{4}$ & 
\quad $1.85 \times 10^{2}$ & \quad $2.96 \times 10^{2}$ & 
\quad $3.40 \times 10^{3}$ & \quad $1.35 \times 10^{5}$ \\ 
LISA + aLIGO  \quad & \quad $3.74 \times 10^{1}$ &
\quad $1.22 \times 10^{-1}$ & \quad $3.03 \times 10^{1}$ & 
\quad $1.52 \times 10^{-1}$ & \quad $2.54 \times 10^{-1}$ & 
\quad $3.98$ & \quad $1.81 \times 10^{2}$ \\ 
\hline\hline
\end{tabular}
}
\endgroup
\caption{Normalized statistical parameter errors 
$\delta {\boldsymbol {\hat \theta}}$ 
with respect to the total multiband SNR $\rho_{\mathrm {tot}}$ 
for System B (GW150914-like BBH). 
The notation is similar to that of Table~\ref{table:sta-errA},
and they are evaluated at the true values 
$\nu = 0.244898$, $\chi_s = 0.8$, and $\chi_a = 0.1$.
} 
\label{table:sta-errB}
\end{table}

Finally, Table~\ref{table:sta-errC} reports 
$\delta {\boldsymbol {\hat \theta}}$ for System C 
(LISA's ``threshold'' BBH) measured by the multiband GW network   
``B-DECIGO  $+$ LISA'',  normalized to $\rho_{\mathrm {tot}}$. 
This clearly shows that B-DECIGO can measure 
${\boldsymbol {\theta}}_{\mathrm {BBH}}$ very well, except for $\chi_a$. 
We find that in terms of the total statistical errors 
${\boldsymbol {\sigma}}_{\mathrm {BBH}}$ of Eq.~\eqref{hat-theta}, 
mass and spin parameters will be determined within 
$\sigma_{\ln m} \sim 1\%$, $\sigma_{\nu}/\nu \sim 2\%$, 
and $\sigma_{\chi_s}/\chi_s \sim 15\%$, respectively, 
while the uncertainty $\sigma_{\chi_a}/\chi_a$ exceeds $100\%$.
However, we also see that earlier LISA analysis will be able to 
(weakly) determine $\nu$ and $\chi_s$ within $\sim 2\%$ 
and $~\sim 70\%$, respectively. 
This will break the degeneracy between $\nu$ and $\chi_{s,a}$ 
that limits the precision of B-DECIGO measurement of $\chi_a$. 
With a joint LISA $+$ B-DECIGO analysis, indeed, 
the cross correlation $c^{\nu \chi_a} \sim 94\%$ in B-DECIGO analysis 
is reduced to $c^{\nu \chi_a} \sim 70\%$, 
and we will be able to get better estimate of $\chi_a$ within $\sim 10\%$. 
Therefore, the joint LISA $+$ B-DECIGO observation 
will be an unique smoking gun to convincingly measure 
intermediate-mass BBHs with spins.

\begin{table}[tb]
\begingroup
\renewcommand{\arraystretch}{1.2}  
\scalebox{0.7}[0.7]{
\begin{tabular}{lc|cccccc}
\hline\hline
Detector  \quad & \quad SNR & 
\quad $\delta {\hat t}_c$ & \quad $\delta {\hat \Psi}_c$ & 
\quad ${\delta {\hat m}} / m$ & \quad ${\delta {\hat \nu}} / \nu$ & 
\quad $\delta {\hat \chi_s} / \chi_s$  & 
\quad $\delta {\hat \chi_a} / \chi_a$ \\
\hline
\multicolumn{8}{l}{B-DECIGO (BD) + LISA}\\
\hline
BD  \quad & \quad $(3.78 \times 10^{2})$ &
\quad $1.43 \times 10^{2}$ & \quad $5.26 \times 10^{2}$ & 
\quad $4.66$ & \quad $7.70$ & 
\quad $5.58 \times 10^{1}$ & \quad $5.77 \times 10^{2}$ \\ 
LISA  \quad & \quad $(3.80 \times 10^{1})$ &
\quad $1.89 \times 10^{3}$ & \quad $2.92 \times 10^{3}$ & 
\quad $4.76$ & \quad $7.91$ & 
\quad $2.67 \times 10^{2}$ & \quad $2.84 \times 10^{3}$ \\ 
BD + LISA  \quad & \quad $3.80 \times 10^{2}$ &
\quad $2.18 \times 10^{1}$ & \quad $4.23 \times 10^{1}$ & 
\quad $4.40 \times 10^{-1}$ & \quad $7.31 \times 10^{-1}$ & 
\quad $4.54 $ & \quad $4.75 \times 10^{1}$ \\ 
\hline\hline
\end{tabular}
}
\endgroup
\caption{Normalized statistical parameter errors 
$\delta {\boldsymbol {\hat \theta}}$ 
with respect to the total multiband SNR $\rho_{\mathrm {tot}}$ 
for System C (LISA's ``threshold'' BBH). 
The notation is similar to that of Table~\ref{table:sta-errA}. 
and they are evaluated at the true values, 
$\nu = 0.138889$, $\chi_s = 0.8$, and $\chi_a = 0.1$.
} 
\label{table:sta-errC}
\end{table}

\subsection{Systematic errors}
\label{subsec:sys_err}

Table~\ref{table:sys-err} reports the mismatch 
($\equiv 1 - {\mathrm{match}}$) with Eq.~\eqref{match} 
and the ratio of the estimate of the systematic parameter errors 
$\Delta {\boldsymbol{\theta}}$ of Eq.~\eqref{sys-err}
to the \textit{overall} statistical errors 
${\boldsymbol {\sigma}}$ of Eq.~\eqref{sta-error}, 
neglecting the $4$PN test-mass term of Eq.~\eqref{F2-4PN} 
in the GW phase $\Psi_{\mathrm {BNS/BBS}}$; 
recall that mismatch and $\Delta {\boldsymbol{\theta}}$ 
are \textit{independent of SNR}. 
For each of the three systems, we see that 
the mismatch always exceeds the $3\%$ mark in higher frequency bands, 
which means that our sky-averaged PN waveform 
${\tilde h}(f)$ in Eq.~\eqref{hf} is \textit{not ``faithful''} 
in the parameter estimation in those bands~\cite{Damour:1997ub}. 
This is expected from the fact that the convergence of PN approximation 
in the late inspiral is likely to be too
slow~\cite{Yunes:2008tw,Sago:2016xsp,Fujita:2017wjq}, 
where we need, e.g., 
effective-one-body formalism~\cite{Bohe:2016gbl} 
or phenomenological models~\cite{Husa:2015iqa,Khan:2015jqa} 
to combine the results from numerical-relativity simulation.
Our conclusion is also in general agreement 
with existing inspiral-only studies 
(e.g., Refs.~\cite{Favata:2013rwa,Yagi:2013baa} for BNSs). 
We shall no longer be concerned with these low-match configurations. 

Surprisingly, we also notice that 
$\Delta {\boldsymbol{\theta}}$ concerning the other high-match cases
always dominates ${\boldsymbol {\sigma}}$ 
unless the mismatch is extremely low (less than $\sim 10^{-3} \%$)
\footnote{Note, however, that for each of the systems considered 
the assumption $\Delta \theta^{a} \partial_a \Psi(f) \lesssim 1$ 
for the formula in Eq.~\eqref{sys-err} is marginally violated 
as the GW frequency $f$ approaches 
the cutoff frequencies $f_{\mathrm {in/end}}$.  
Hence, it is quite likely that $\Delta {\boldsymbol{\theta}}$ 
is overestimated; 
a more sophisticated method will be necessary to reliably 
compute the systematic mismodeling errors~\cite{Cutler:2007mi}.},
which shows that the measured BNS and BBH parameters 
resulting from ${\tilde h}(f)$ are strongly biased 
by $4$PN test-mass phase terms. 
Essentially, BNS and BBH inspirals in lower frequency bands 
can be high-SNR sources, greatly reducing ${\boldsymbol {\sigma}}$.

\begin{table}[tb] 
\scalebox{0.7}[0.7]{
\begin{tabular}{lc|cccccc}
\hline\hline
Detector & mismatch $(\%)$ &
\quad $|\Delta t_c / \sigma_{t_c}|$  & 
\quad $|\Delta \Psi_c / \sigma_{\Psi_c}|$ & 
\quad $|{\Delta m} / \sigma_{m}|$ & 
\quad $|{\Delta \nu} / \sigma_{\nu}|$ & 
\quad $|\Delta \chi_s / \sigma_{\chi_s}|$  & 
\quad $|\Delta \chi_a / \sigma_{\chi_a}|$ \\
\hline
\hline
\multicolumn{8}{l}{System A: GW170817-like BNS}\\
\hline
aLIGO  \quad &  \quad  $3.52$ &
\quad $\cdots$ & \quad $\cdots$ & 
\quad $\cdots$ & \quad $\cdots$ & 
\quad $\cdots$ & \quad $\cdots$ \\ 
ET  \quad &  \quad  $2.94$ &
\quad $\cdots$ & \quad $\cdots$ & 
\quad $\cdots$ & \quad $\cdots$ & 
\quad $\cdots$ & \quad $\cdots$ \\ 
B-DECIGO  \quad &  \quad  $4.96 \times 10^{-3}$ &
\quad $4.80 \times 10^{-4}$ & \quad $5.87 \times 10^{-3}$ & 
\quad $2.47 \times 10^{-6}$ & \quad $5.84 \times 10^{-8}$ & 
\quad $2.86 \times 10^{-4}$ & \quad $2.85 \times 10^{-4}$ \\ 
\hline
\multicolumn{7}{l}{System B: GW150914-like BBH}\\
\hline
aLIGO  \quad &  \quad  $5.14$ &
\quad $\cdots$ & \quad $\cdots$ & 
\quad $\cdots$ & \quad $\cdots$ & 
\quad $\cdots$ & \quad $\cdots$ \\ 
ET  \quad &  \quad  $7.15$ &
\quad $\cdots$ & \quad $\cdots$ & 
\quad $\cdots$ & \quad $\cdots$ & 
\quad $\cdots$ & \quad $\cdots$ \\ 
B-DECIGO  \quad & \quad $9.02 \times 10^{-1}$ &
\quad $4.39 \times 10^{1}$  & \quad $2.83 \times 10^{1}$ & 
\quad $1.30 \times 10^{1}$ & \quad $1.30 \times 10^{1}$ & 
\quad $2.27 \times 10^{1}$ & \quad $2.27 \times 10^{1}$ \\ 
LISA  \quad & \quad $1.16 \times 10^{-4}$ &
\quad $4.28 \times 10^{-3}$  & \quad $9.06 \times 10^{-4}$ & 
\quad $4.65 \times 10^{-4}$ & \quad $4.65 \times 10^{-4}$ & 
\quad $7.39 \times 10^{-4}$  & \quad $7.40 \times 10^{-4}$ \\ 
\hline
\multicolumn{7}{l}{System C: LISA's threshold BBH}\\
\hline
B-DECIGO  \quad &  \quad  $21.7$ &
\quad $\cdots$ & \quad $\cdots$ & 
\quad $\cdots$ & \quad $\cdots$ & 
\quad $\cdots$ & \quad $\cdots$ \\ 
LISA  \quad & \quad $3.54 \times 10^{-1}$ &
\quad $1.75$     & \quad $1.40$ & 
\quad $6.68 \times 10^{-1}$ & \quad $6.68 \times 10^{-1}$ & 
\quad $1.11$ & \quad $1.11$ \\ 
\hline\hline
\end{tabular}
}
\caption{aLIGO, ET, B-DECIGO, and LISA mismatch 
as well as the ratio of the systematic errors 
$\Delta {\boldsymbol{\theta}}$ 
to the overall statistical errors ${\boldsymbol {\sigma}}$, 
when not including the $4$PN test-mass term of Eq.~\eqref{F2-4PN} 
in the GW phase. 
In each column, the blank cells indicate that systematic errors are too large 
($\Delta {\boldsymbol{\theta}}/{\boldsymbol{\theta}} > 100\%$) 
due to $\text{mismatch} \gtrsim 3\%$. 
For System A observed by B-DECIGO, 
the ratios for the quadrupole and tidal parameters are 
$|\Delta {\tilde {\mathcal {Q}}} / \sigma_{\tilde {\mathcal {Q}}}|
= 2.54 \times 10^{-4}$ and 
$|\Delta {{\tilde \Lambda}} / \sigma_{\tilde \Lambda}| 
= 9.91 \times 10^{-4}$, respectively: nevertheless, 
we should recall that B-DECIGO statistical errors are 
$\sigma_{\tilde {\mathcal {Q}}} / {\tilde {\mathcal {Q}}}> 100\%$ 
and $\sigma_{\tilde {\Lambda}} / {\tilde {\Lambda}}> 100\%$, respectively.}
\label{table:sys-err}
\end{table}

We note that for all of Systems A, B, and C, 
the inclusion of NS or BH spin effects to the GW phase is highly recommended 
although NS spins in waveform modeling are often neglected 
as they are much smaller than BH spins.  
Making use of the formula in Eq.~\eqref{match}, for example, 
the mismatch with and without the highest spinning point-particle terms 
at the $3.5$PN order (Eq.~(6b) of Ref.~\cite{Mishra:2016whh})
are all above the $3\%$ mark when the body's spin is $\chi_i \gtrsim 0.2$; 
this value of spin is essentially independent of the choice of detectors, 
and becomes much smaller when not including the lower-order PN spin terms 
for the limit of $3\%$ mismatch. 

These results motivate the continued development of PN waveform modeling 
for accurate parameter extraction from future B-DECIGO and LISA measurement. 

\section{Discussion}
\label{sec:discuss}

For each system considered in this work, 
we have shown that multiband measurement with B-DECIGO 
will determine the binary parameters 
(masses, spins, NS quadrupole parameters and Love numbers) 
with percent-level precision, 
if the systematic bias due to PN waveform mismodeling is under control. 
Consequently, this exquisite precision will be able to enhance 
the already (or expected) rich payouts gained from aLIGO and LISA observation. 
Below we shall highlight two potential examples 
from cosmography and fundamental physics, 
focusing on the joint B-DECIGO $+$ aLIGO (ET) measurement 
of the GW170817-like BNS and GW150914-like BBH, respectively.

The first example is 
\textit{the redshift measurement of GW170817-like BNS 
using only GW observation}, 
proposed by Messenger and Read~\cite{Messenger:2011gi} 
as well as Harry and Hinderer~\cite{Harry:2018hke}. 
Assuming that the NS equation of state is well constrained 
from the BNS waveform of the late inspiral and merger~\cite{Dietrich:2018uni}, 
the measurement of NS quadrupole parameters $\kappa_i$ and 
tidal deformabilities ${\Lambda}_i$ $(i = 1, 2)$ imprinted 
in the (early) inspiral waveform allows determination 
of the source redshift $z$ directly. 
Essentially, the point is that the parameters $(\kappa_i,\,{\Lambda}_i)$
manifestly depend on the source-frame masses $m_i^S$ scaling as 
${\kappa}_i \sim (m_i^S)^{-3} \sim (m_i^O)^{-3} (1 + z)^{3}$  
and 
${\Lambda}_i \sim (m_i^S)^{-5} \sim (m_i^O)^{-5} (1 + z)^{5}$, 
respectively, where $m_i^O$ are the observer-frame NS masses. 
Because $(\kappa_i,\,{\Lambda}_i)$ are related with $m_i^S$ 
through the NS equation of state, 
the measurement of $(\kappa_i,\,{\Lambda}_i)$ and $m_i^O$ 
is then translated to $m_i^S$ and $m_i^O$, 
from which we can determine the source redshift $z$. 

This approach to cosmography benefits 
from the joint B-DECIGO $+$ ET measurement. 
The Einstein Telescope will place strong constraints 
on the NS spins and tidal influence in BNSs,  
while B-DECIGO can precisely measure the NS masses: 
recall Table~\ref{table:sta-errA}.
Following the method of Ref.~\cite{Messenger:2011gi} 
to examine the uncertainties in our inspiral-only waveform parameters, 
we estimate $\boldsymbol{\theta}_{\mathrm {BNS}}$ in Eq.~\eqref{theta}
for System A relocated at $\sim 2.92 \,{\mathrm {Gpc}}$ ($z = 0.50$), 
leaving $({\tilde {\mathcal Q}},\,{\tilde \Lambda})$ out 
but with the addition of $z$, making use of Eqs.~\eqref{sta-error}.
The statistical errors for $z$ then read 
\begin{equation}
\delta z / z
=
\left\{
(1.17 \times 10^{-1})_{\text {ET}}
,\,
(6.93 \times 10^{-2})_{\text {ET+BD}}
\right\}\,,
\end{equation}
where the SNRs are $\rho_{\mathrm {ave}}^{\mathrm {ET}} = 6.62$
and $\rho_{\mathrm {tot}} = 7.00$. 
Notably, we see that the uncertainty is $\sim 60\%$ 
of what would be obtained with ET analysis alone. 
Joint B-DECIGO $+$ ET measurement of BNSs 
would thus be quite valuable for this cosmography.


The second example is \textit{constraining 
the mass and spin of the final remnant BH} after merger. 
For non-precessing BBHs, there are now numbers 
of numerical-relativity fitting formulas 
that enables the mapping of the initial BH masses $m_{1,2}$ and 
dimensionless spin parameters $\chi_{1,2}$ 
to the final mass $M_f = M_f (m,\, \nu,\, \chi_s,\, \chi_a) $ 
and dimensionless spin parameter 
$\chi_f \equiv |{\bf S}_f|/M_f^2 = \chi_f(m,\, \nu,\, \chi_s,\, \chi_a)$ 
of the remnant Kerr
BH~\cite{Healy:2014yta,Healy:2016lce,Jimenez-Forteza:2016oae,Healy:2018swt}. 
Because B-DECIGO (and ET) will measure $m_{1,2}$ and $\chi_{1,2}$ 
very precisely, it will set stringent constraints on $M_f$ and $\chi_f$, 
making use of these fitting formulas.

We estimate the precision with which $M_f$ and $\chi_f$ 
for System B can be determined from the inspiral phase, 
adopting the fitting formulas (``UIB formulas'')
~\cite{Jimenez-Forteza:2016oae} available 
in LALInference~\cite{Veitch:2014wba, lalsuite}. 
Using the statistical errors 
${\delta {\boldsymbol {\hat \theta}}}_{\mathrm {BBH}}$ 
in Table~\ref{table:sta-errB} as input, 
the corresponding statistical errors on $M_f$ and $\chi_f$ are given 
by a standard variance propagation of non-linear functions:
see footnote 19.
Here, we neglect the systematic bias in the fitting formulas 
(see Ref.~\cite{Ghosh:2017gfp} for details). 
After simple algebra, we obtain $(M_f/m, \chi_f) = (0.891, 0.900)$ and 
\begin{equation}
\left({\delta M_f} / M_f,\, {\delta \chi_f} / \chi_f \right)
=
\left\{
(1.20 \times 10^{-3},\, 3.86 \times 10^{-3})_{\text{BD}} ,\, 
(7.47 \times 10^{-4},\, 2.57 \times 10^{-3})_{\text{BD+aLIGO}} 
\right\}\,.
\end{equation}
The joint B-DECIGO $+$ ET analysis will 
further reduce the uncertainties in $M_f$ and $\chi_f$ 
by up to factor of four in B-DECIGO $+$ aLIGO analysis. 
Given that aLIGO and ET independently measure 
$M_f$ and $\chi_f$ from the merger-ringdown phase, 
this drastic improvement in estimation from the inspiral phase
will help to strengthen the inspiral-merger-ringdown consistency test 
of general relativity~\cite{Hughes:2004vw,Nakano:2015uja,
Ghosh:2016qgn,Ghosh:2017gfp,Cabero:2017avf}. 
Other extreme gravity tests that can be done with binary inspirals, 
including direct and parameterized test of gravity
theory~\cite{Gair:2012nm,Berti:2015itd,Berti:2018cxi} 
and the ``no-hair'' test of BBH nature~\cite{Krishnendu:2017shb}, 
might benefit from B-DECIGO and multiband measurement as well. 


In conclusion, we expect B-DECIGO measurement of BNSs and BBHs 
in the deci-Hz band will complement, e.g., aLIGO, advanced Virgo, KAGRA, 
IndIGO, ET and LISA observation in the hecto-Hz and milli-Hz bands, 
boosting our understanding of astrophysics, cosmology, and gravity science. 
In the era of multiband GW astronomy, we shall decide and go, 
DECIGO~\cite{Seto:2001qf,Nakamura:2016hna}. 

\section*{Acknowledgments}
We thank R. Sturani and all participants of ``DECIGO Workshop 2017'' 
for useful discussion.
SI acknowledges financial support 
from Ministry of Education - MEC during his stay at IIP-Natal-Brazil. 
This work was supported in part by JSPS/MEXT KAKENHI Grant No.\ JP16K05347, 
No.\ JP17H06358 (H.N.) and No.\ JP15H02087 (T.N.). 
All the analytical and numerical calculations in this paper 
have been performed with \textit{Maple}.



\end{document}